\documentclass[twocolumn,secnumarabic,amssymb,amsfonts, reprint, nobibnotes, aps, pra]{revtex4-2}
\usepackage{bm}
\usepackage[T1]{fontenc}
\usepackage{amsthm}
\usepackage{mathtools}
\usepackage{physics}
\usepackage{xcolor}
\usepackage{graphicx}
\usepackage[left=23mm,right=13mm,top=35mm,columnsep=15pt]{geometry} 
\usepackage{adjustbox}
\usepackage{placeins}
\usepackage[T1]{fontenc}
\usepackage{lipsum}
\usepackage{csquotes}

\usepackage{subcaption}
\begin{document}
	\title{Tracking nanoscale perturbation in active disordered media}
\author{Renu Yadav$^1$, Patrick Sebbah$^2$, Maruthi M. Brundavanam$^1$, and Shivakiran Bhaktha B. N. $^1$} 	\email[Correspondence email address: ]{kiranbhaktha@phy.iitkgp.ac.in}
\affiliation{$^1$Department of Physics, Indian Institute of Technology Kharagpur, Kharagpur-721302, India \\ $^2$Department of Physics, The Jack and Pearl Resnick Institute for Advanced Technology,
	Bar-Ilan University, Ramat-Gan, 5290002 Israel }
	\begin{abstract}
	The disorder induced feedback makes random lasers very susceptible to any changes in the scattering medium. The sensitivity of the lasing modes to perturbations in the disordered systems have been utilized to map the regions of perturbation. A tracking parameter, that takes into account the cumulative effect of changes in the spatial distribution of the lasing modes of the system has been defined to locate the region in which a scatterer is displaced by a few nanometers. We show numerically that the precision of the method increases with the number of modes. The proposed method opens up the possibility of application of random lasers as a tool for monitoring locations of nanoscale displacement which can be useful for single particle detection and monitoring. 
	\end{abstract}
	
	
	\maketitle
	\section{Introduction}
	A random laser (RL) is an optical device that utilizes the disorder in the system for the optical feedback. Unlike conventional lasers, no well-defined cavities are present in RLs. The idea of feedback by multiple scattering was first proposed by Letokhov \cite{letokhov1968optical} and has been extensively used to realize random lasing in a variety of disordered systems \cite{tulek2010studies,lu2015random,cao1999random,song2010random,lawandy1994laser,shivakiran2012optofluidic,ferjani2008random}. Two types of RLs have been reported namely, coherent RLs and incoherent RLs, depending on whether the scattering induces the feedback in the field or the intensity, respectively \cite{cao2003lasing}. The scattering strength determines the lasing characteristics such as the lasing threshold of the system, spatial confinement of the modes, etc. Based on the scattering strength, disordered systems can be broadly divided into two categories, namely, strongly scattering and weakly scattering systems. In the strongly scattering systems the lasing modes are localized well within the system and are identical to the quasi-bound (QB) states of the passive system \cite{sebbah2002random,jiang2002localized,andreasen2011modes}, whereas, in weakly scattering systems the lasing modes extend all over the system \cite{andreasen2011modes,vanneste2007lasing}.
	
	Unlike conventional lasers, RL emission is random in wavelength, omnidirectional \cite{cao1999random} and has low spatial and temporal coherence \cite{gouedard1993generation,noginov1999interferometric,redding2011spatial}. These properties make them suitable for different applications like, imaging \cite{redding2012speckle}, displays and lighting \cite{chang2018white}, holography \cite{mallick2020holographic}, etc., but it limits their use where specific wavelength or unidirectional emission is required. Spatial light modulators (SLMs) have been used to shape the pump intensity profile to control the emission and directionality of RLs making them useful for different applications \cite{bachelard2012taming,hisch2013pump,liew2014active,liew2015pump,ge2015selective,bachelard2014adaptive,kumar2021localized}. As the feedback in RLs is provided by disorder-induced scattering, the lasing modes are very sensitive to any changes in the scattering medium. This makes RLs a natural candidate for designing sensors for various applications. The strong dependence of emission characteristics of RLs on the scattering properties of the medium have been utilized to assess nanoscale perturbations \cite{ho2012random}. The monitoring of single nanoparticle perturbation enables to detect single virus, bacterium and biolmolecule. Random lasers have been used as a diagnostic tool for bio-imaging and bio sensing in various biological structures infiltrated with dye \cite{polson2004random,song2010random,siddique1995mirrorless}. The nanoscale deformation and prefailure damage in bones can be detected by monitoring the shifts in the random lasing peaks \cite{song2010detection}. In ex-vivo dye infiltrated human tissues, the changes in the emission spectrum have been observed in malignant tissues as compared to the healthy ones \cite{polson2010cancerous}. The cancerous tissues of different grades of malignancy can be differentiated as they exhibit different lasing spectra for same pump energy \cite{wang2017random}. RLs have been proposed as an in-vivo tool to differentiate between skin, fat, muscle and nerve tissues during laser surgery \cite{hohmann2019investigation}. 
	
	In this work, RLs have been proposed as a tool to map the regions of nanoscale perturbation in several random media. A two dimensional (2D) active disordered system has been considered and nanoscale perturbations have been introduced in the medium. Using finite difference time domain (FDTD) method \cite{taflove1995finite} the modes and the corresponding spatial field distributions for the system before and after the perturbation have been computed. In the past, RLs have been used to detect changes in the scattering medium \cite{ho2012random}. In this work we go a step further and show numerically that it is also possible to identify the position of the perturbation with good precision. A small perturbation in the system leads to minute changes in the spectral position of the modes and their corresponding spatial field distributions, but the individual modes do not provide any information about the location of the perturbation. So, a tracking parameter is defined which takes into account the cumulative effect of changes in the modes, to map the region of perturbation. We find that its mapping converges to the defect location when the number of modes increases. This finding paves the way to single particle tracking in disordered systems. The theoretical explorations in this work provide an initial framework to utilize RLs in the field of diagnostics to monitor and track the growth of tumors in disordered biological systems.
	
	\section{Numerical method and computational details}
	A 2D disordered system of size, $L^{2}= 5 \times 5 $ $\mu m^2$ has been considered. It consists of circular particles with radius, $r =60 \hspace{0.1 cm} nm$ and refractive index,  $n_2=2.54$, randomly distributed in a background medium of refractive index, $n_1=1.53$. The values of the refractive index have been chosen to mimic the presence of $TiO_2$ particles in 4-(Dicyanomethylene)-2-methyl-6-(4-dimethylaminostyryl)-4H-pyran (DCM) doped polyvinyl alcohol (PVA) thin films \cite{sarkar2017effect,choubey2020origin,choubey2020random}. The background medium has been chosen as the active part of the system and modeled as a four level atomic system. The surface filling fraction of the scatterers is $28\%$. 
	In this study, 2D FDTD computation has been carried out using transverse magnetic fields with a grid resolution of $\Delta x= \Delta y=10 \hspace{0.1 cm}nm$, along $x$ and $y$ directions, respectively. In order to ensure the stability of the simulation, the time step  chosen is, $\Delta t =2.37 \times 10^{-17} s$ \cite{yee1966numerical}. The parameters used for the active medium are mentioned in Ref. \cite{sebbah2002random}. The system is pumped uniformly with a Gaussian pulse of central wavelength $ 532 \hspace{0.1 cm}nm$ and pulse duration  $ \sim 10^{-15}$ s at a pump level above the lasing threshold of system.
	
	\section{Results and discussion}
	
\begin{figure}[htbp]
		\includegraphics[width=1\linewidth, height=4.5 cm]{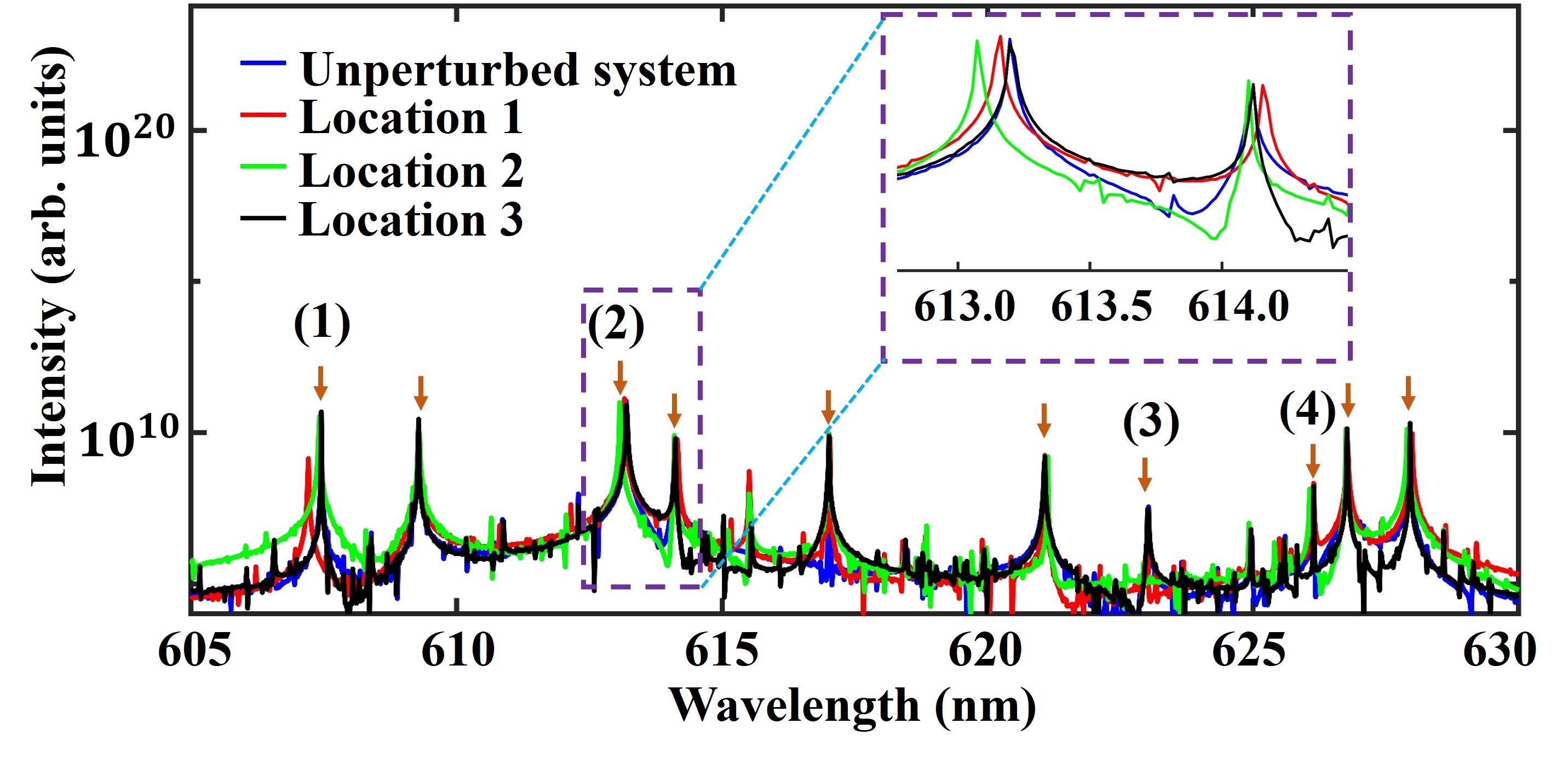}
		\captionsetup{justification=raggedright}
		\caption{Emission spectra of the unperturbed system  and the system with a single particle perturbed by $ 10\hspace{0.1 cm} nm$ at three different locations in the system. Ten peaks considered are marked with arrows. The labeled modes are  $607.45$ nm (1), $613.20$ nm (2), $623.01$ nm (3), and $626.13$ nm (4). The inset shows the magnified image of the region marked with dashed magenta line. It shows the spectral shift of the mode as the perturbation is introduced in the system. }
		\label{fig3}
	\end{figure}

\begin{figure} 
		\begin{subfigure}[b]{4cm}
			\centering
			\includegraphics[width=4cm]{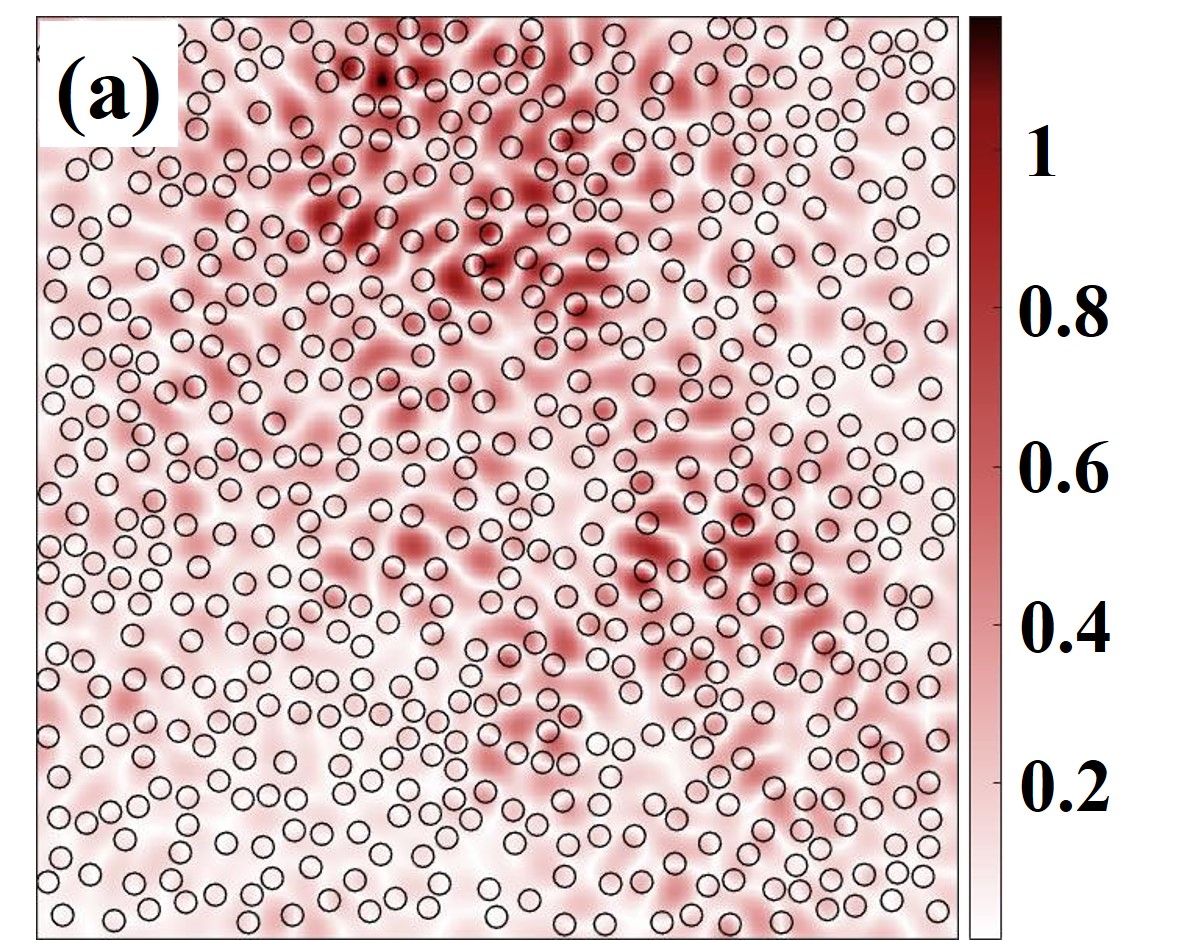} 
			\label{fig2:a} 
		\end{subfigure}
		\begin{subfigure}[b]{4cm}
			\centering
			\includegraphics[width=4cm]{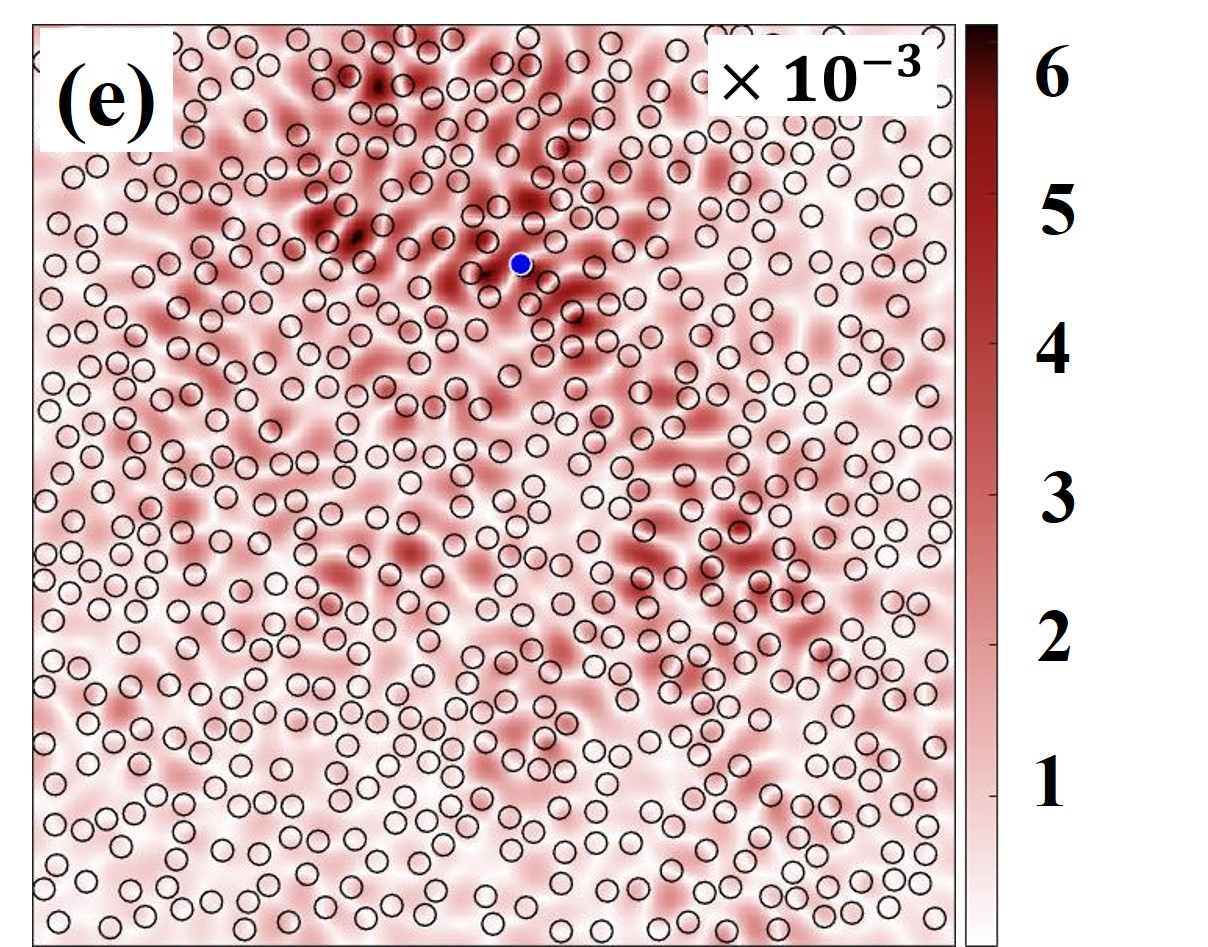} 
			\label{fig2:e} 
		\end{subfigure} 
		\\[-3ex]
		\begin{subfigure}[b]{4cm}
			\centering
			\includegraphics[width=4cm]{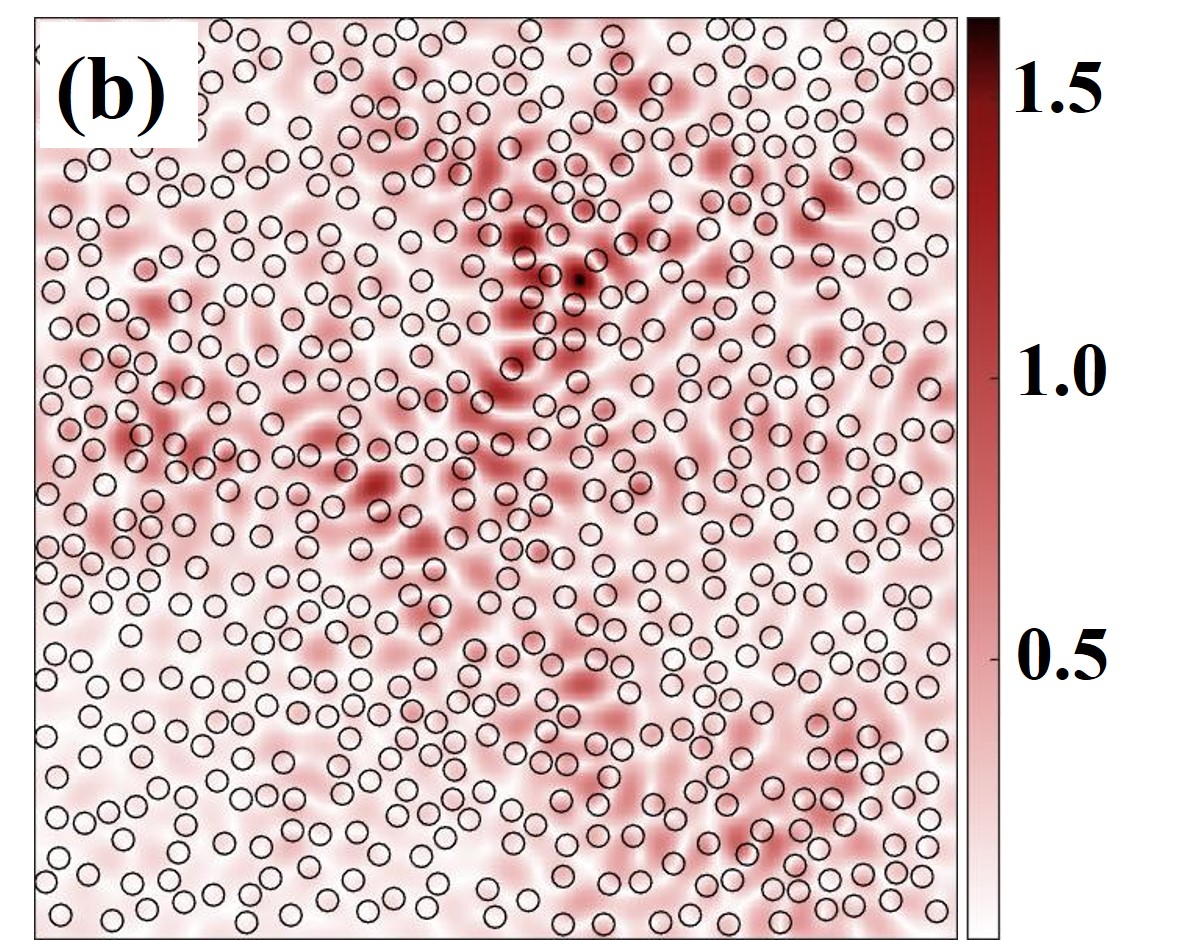} 
			\label{fig2:b} 
		\end{subfigure}
		\begin{subfigure}[b]{4cm}
			\centering
			\includegraphics[width=4cm]{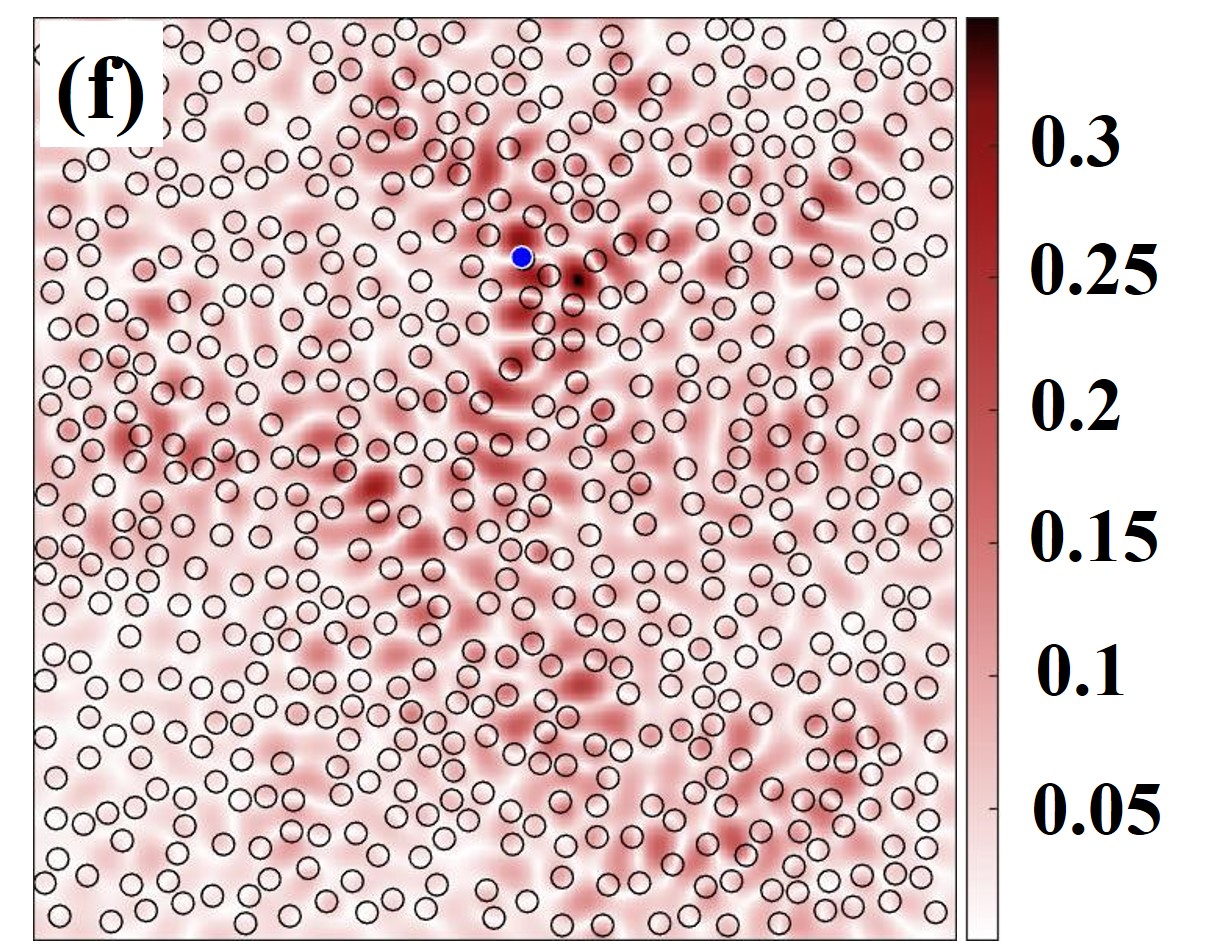} 
			\label{fig2:f} 
		\end{subfigure} 
		\\[-3ex]
		\begin{subfigure}[b]{4.05cm}
			\centering
			\includegraphics[width=4.05cm]{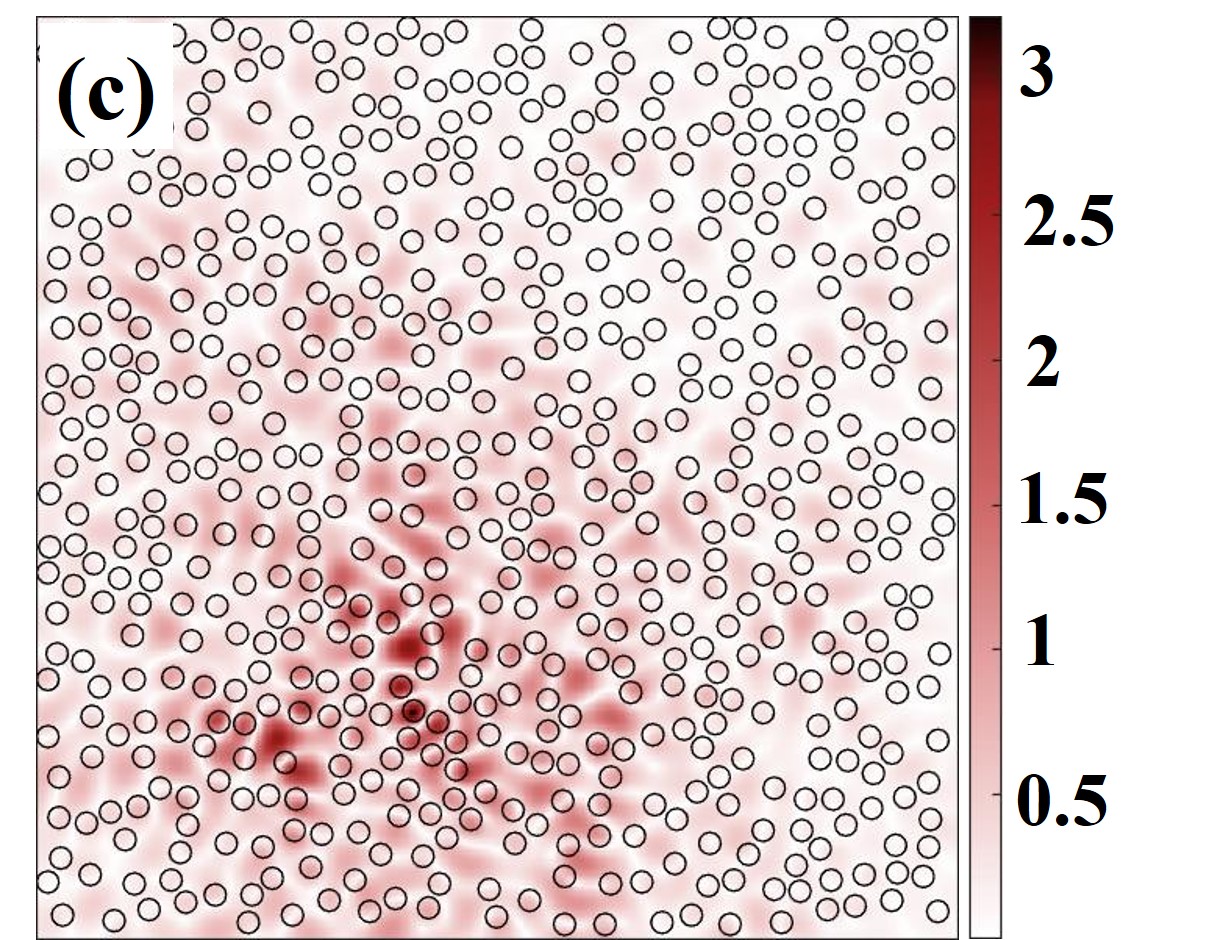} 
			\label{fig2:c} 
		\end{subfigure}
		\begin{subfigure}[b]{4cm}
			\centering
			\includegraphics[width=4cm]{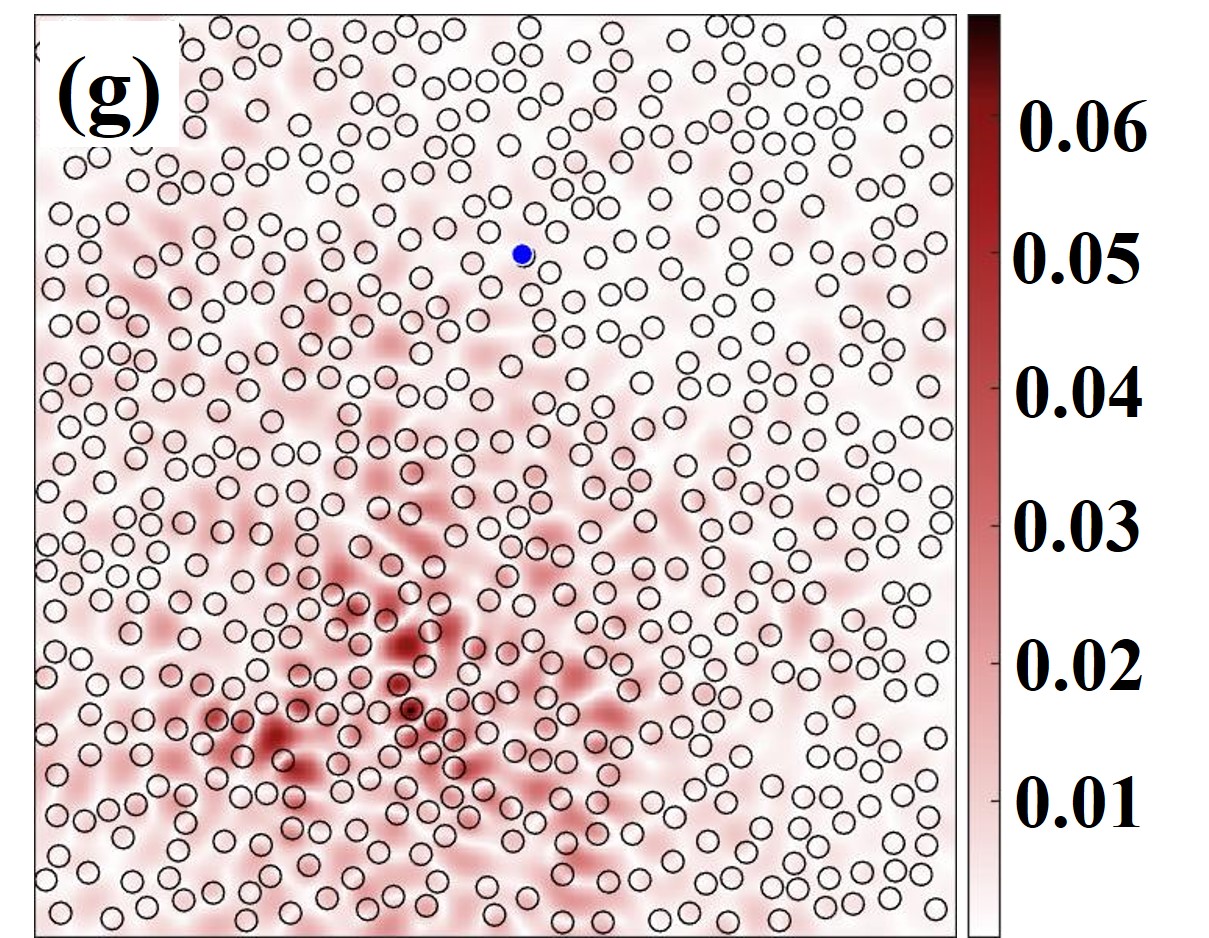} 
			\label{fig2:g} 
		\end{subfigure} 
		\\[-3ex]
		\begin{subfigure}[b]{4cm}
			\centering
			\includegraphics[width=4cm]{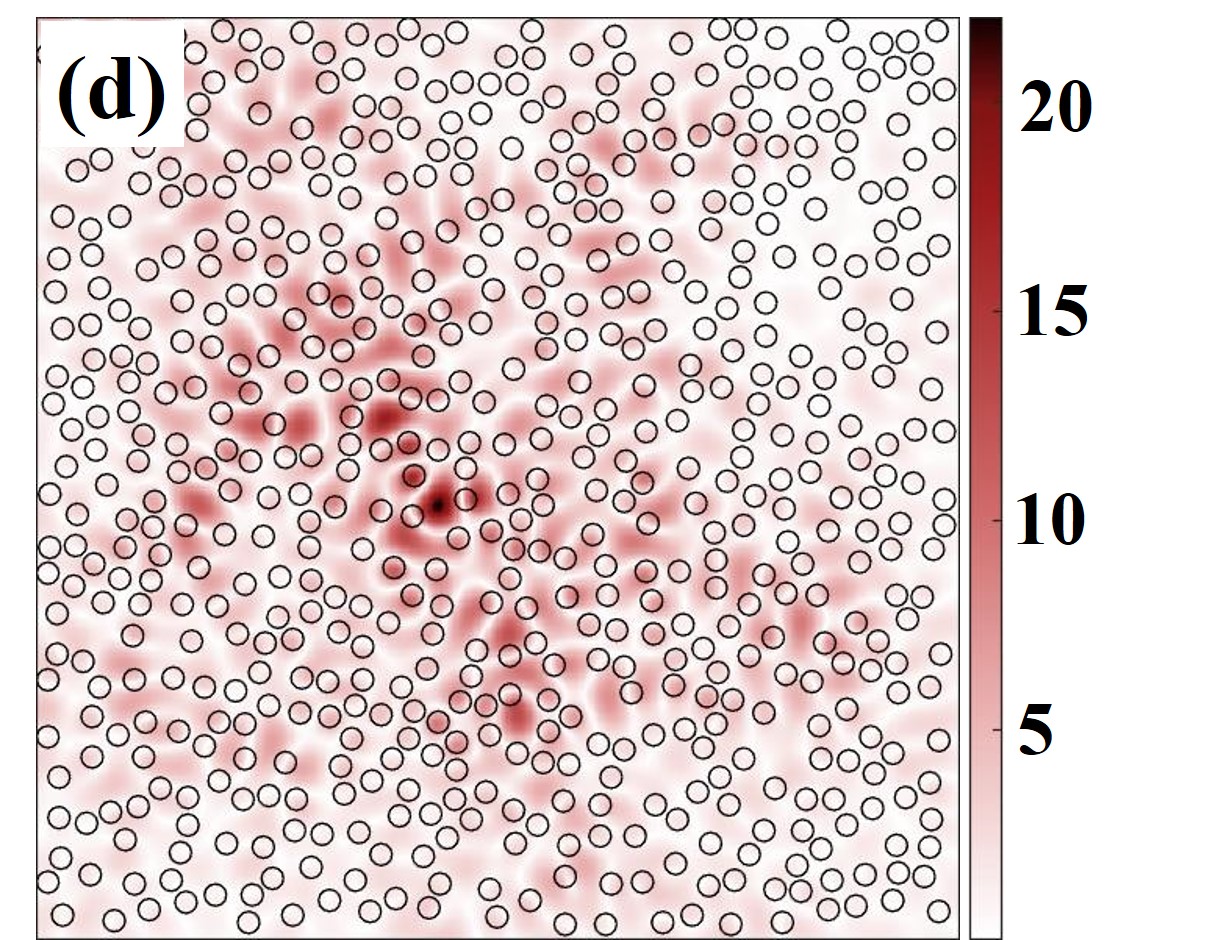} 
			\label{fig2:d} 
		\end{subfigure}
		\begin{subfigure}[b]{4cm}
			\centering
			\includegraphics[width=4cm]{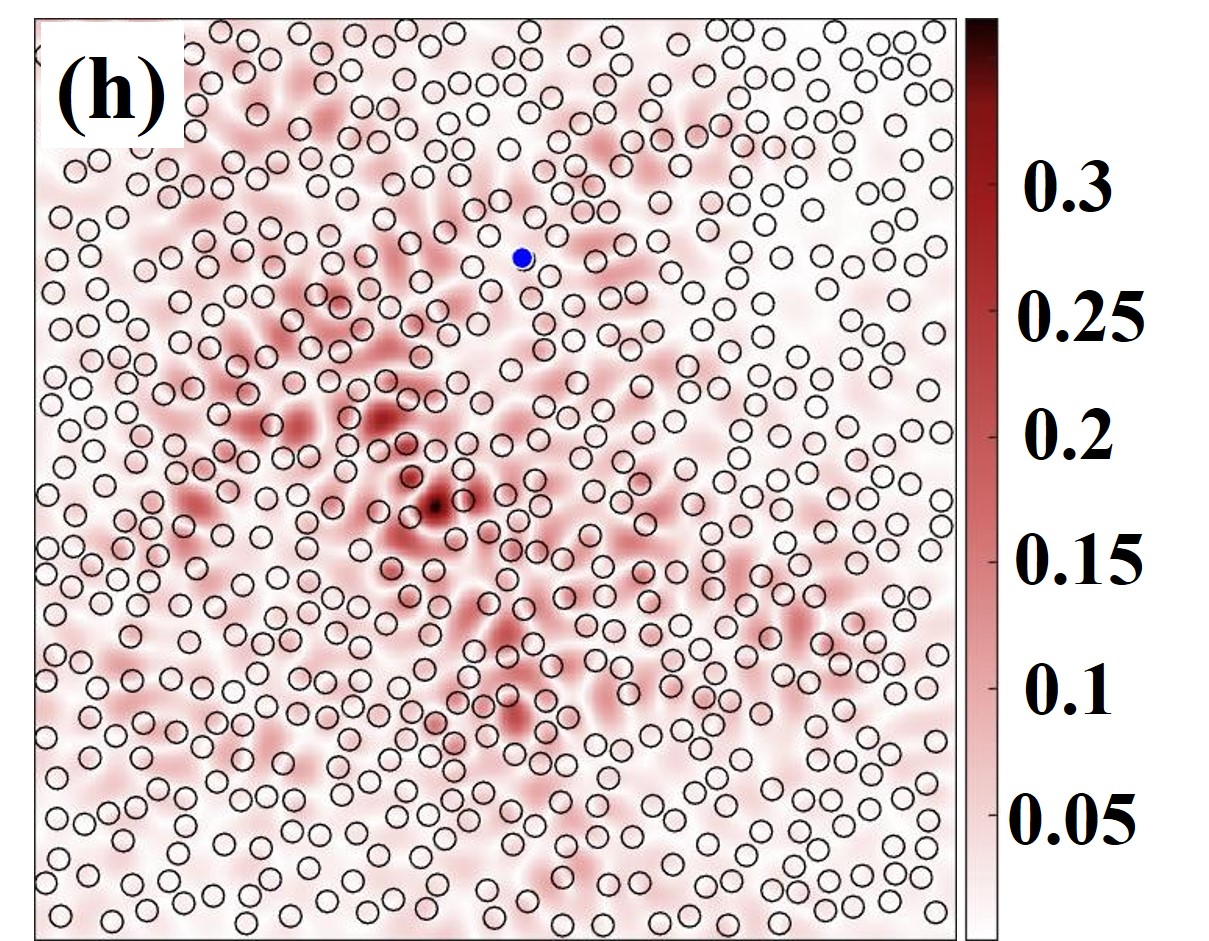} 
			\label{fig2:h} 
		\end{subfigure} 
		\captionsetup{justification=raggedright}
		\caption{Spatial field distribution of modes marked as (1-4) in Fig. \ref{fig3}, before (a-d), and after (e-h) a single particle is displaced by $ 10 \hspace{0.1 cm}nm$. The perturbed particle is marked in blue.}
		\label{fig2} 
	\end{figure}

	The 2D active, random system was pumped above the lasing threshold. The energy in the system was observed to grow exponentially, and after some strong relaxation oscillations, it eventually reaches a steady-state. The lasing modes of the system are calculated by the Fourier transform of the time records of the field after the system has reached the stationary state. Several distinct peaks are observed as shown in Fig. \ref{fig3}. The ten modes considered for further analysis are marked with arrows. The discrete peaks in the emission spectrum indicate lasing action with resonant feedback. The spatial field distribution of the modes is computed by taking the Fourier transform of the field recorded at each grid point. The spatial field distribution of the modes marked as (1-4) in Fig. \ref{fig3} is shown in Figs. \ref{fig2}(a-d). It is observed that the modes are confined well within the system, indicating that the system is strongly scattering. The numerically computed scattering mean free path for the system using Mie scattering theory is, $l_s \approx 0.3 \hspace{0.1 cm}\mu m$ \cite{hulst1981light}. The localization length for the system is calculated by considering the field intensity profiles of modes averaged along $x$ or $y$ directions. The averaged intensity profile exhibits strong local fluctuations but its envelope decays exponentially whose characteristic length, gives the localization length of the modes. The average localization length calculated for the system is, $\xi \approx 2.6 \hspace{0.1 cm}\mu m$. The scattering mean free path and the localization length also indicate that system is strongly scattering and the modes are confined well within the system, respectively.
	
	Next, in order to introduce a single nanoscale perturbation in the system, a randomly chosen scatterer was displaced by $10 \hspace{0.1 cm}nm$ along an arbitrary direction. The numerical parameters limit the minimum and the maximum perturbation that can be introduced in the system. The minimum displacement cannot be smaller than the grid resolution and  maximum displacement possible is dependent on the surface filling fraction of the system. In realistic systems, such limitations don't exist and hence it is expected that even smaller perturbations can be detected. The effect of perturbation at different locations in the system, on the RL spectra and spatial field distribution of modes has been studied. The emission spectra of the system, when a single particle is perturbed at three different locations in the system are shown in Fig. \ref{fig3}. It is observed that the perturbation causes slight shift in the spectral positions of the random lasing modes, as shown in the inset. The modes 1-4 experience spectral shifts ($\Delta \lambda$) of magnitude $0.239 \hspace{0.1 cm} nm$, $0.039 \hspace{0.1 cm} nm$, $0.004 \hspace{0.1 cm} nm$ and $0.004 \hspace{0.1 cm} nm$, respectively. Further details on the correspondence between the magnitude of the spectral shifts and the mode field distributions are presented later. 
	
\begin{figure}[t]
		\centering
		\includegraphics[width=8 cm, height=6 cm]{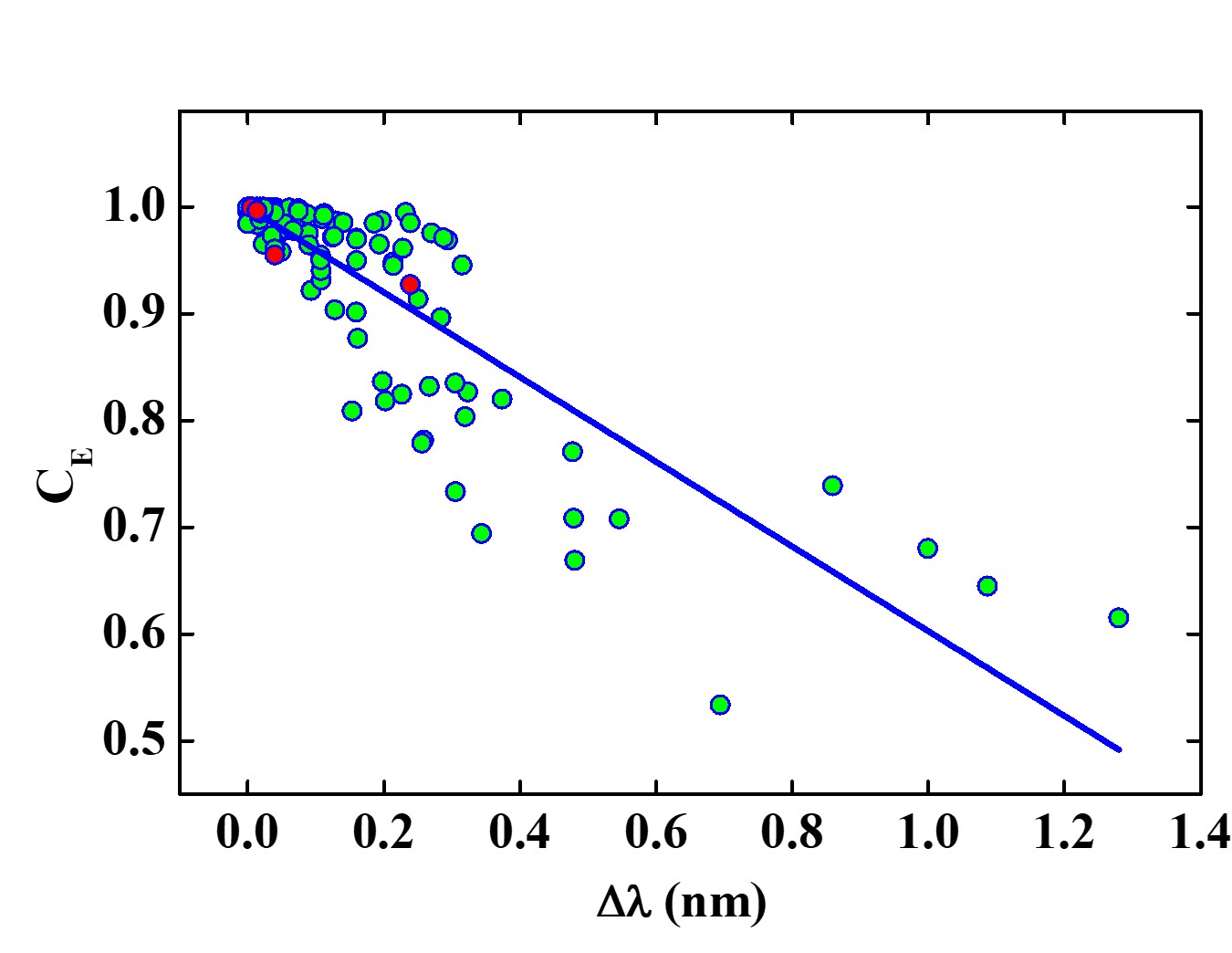}
		\caption{Variation of $C_E$ with the spectral shift ($\Delta \lambda$) in the lasing modes due to perturbation introduced in the system. The solid line represents the linear fit to the data points. The data points marked in red correspond to modes 1-4 in Fig. \ref{fig2}.}
		\label{lcefig}
	\end{figure}

	The perturbation also leads to changes in the spatial field distribution of modes as shown in Figs. \ref{fig2}(e-h), for the perturbation at location 1 (Fig. \ref{fig3}), wherein the perturbed particle is marked in blue. In order to quantify the changes in the system due to perturbation, 2D correlation coefficient ($C_E$) between the spatial field distribution of modes before and after perturbation is calculated, which is defined as:
	\begin{equation} \label{1}
		C_E= \frac{\sum_{x} \sum_{y}(E(x,y)-\bar E)(E^{'}(x,y)-\bar E^{'})}{\sqrt{\left ( \sum_{x} \sum_{y}(E(x,y)-\bar E)^2 \right ) \left ( \sum_{x} \sum_{y} (E^{'}(x,y)-\bar E^{'})^2 \right )}}
	\end{equation}
	where, $E(x,y)$ and $E^{'}(x,y)$  are the field magnitudes of the modes at location $(x,y)$ in the system, before and after perturbation, respectively. $\bar{E}$ and $\bar E^{'} $ represent the mean field values of the corresponding modes. The $C_E$ value quantifies the similarity between the modes before and after the perturbation, and for the system shown in Fig. \ref{fig2}, it is found to be 0.92, 0.95, 0.99, and 0.99 for modes 1, 2, 3 and 4, respectively. The $C_E$ values indicate that the perturbation leads to more changes in modes 1 and 2 as compared to modes 3 and 4. It is evident from Figs. \ref{fig2}(e-f) that the perturbed scatterer is present in the region with high field value for modes 1 and 2 as compared to modes 3 and 4. Thus, a perturbation in the high field region of a mode leads to more changes in the lasing modes as compared to a perturbation in low field region. It is also observed that the shape of the spatial field profiles of modes do not change drastically after the perturbation. However, minute changes are observed in the distribution of the field and its magnitude. Moreover, the perturbation also leads to changes in the spectral location of the lasing modes. The spectral shift ($\Delta \lambda$) in the modes is linearly related to $C_E$ value as shown in Fig. \ref{lcefig}. For modes exhibiting small changes in their spectral position, the $C_E$ values are found to be $\sim$ 1, and as $\Delta \lambda$ increases, $C_E$ decreases. Thus, with increasing spectral shift the changes in the field distribution of the corresponding modes become more prominent.

	The sensitivity of lasing modes to nanoscale displacements has been utilized to monitor perturbations in the system. It is observed that the nanoscale alteration in the scatterer position leads to changes in the lasing modes and their spatial field distributions, and the amount of change varies for each mode. But, these changes in the individual modes do not provide any information about the position of the perturbed particle. Now, it is interesting to ask a question on whether one can identify the particle that has been perturbed, given the modes before and after the perturbation are known. Here, we show that it is possible to locate the position of the scatterer that has been perturbed with the help of the computed modes by defining a tracking parameter, TP as,
	\begin{equation} \label{2}
		TP(x,y)= \prod_{m} \lvert E_{m}(x,y)-E_{m}^{'}(x,y) \rvert
	\end{equation}
	where $E_{m}(x,y)$ and $E_{m}^{'}(x,y)$ are the normalized field values of mode $m$ before and after the perturbation at $(x,y)$ position in the system, respectively. The tracking parameter is given by the product of change in the field distribution of the modes considered, due to perturbation. Here, the normalized field values have been considered as we are interested in how the field at a point in the system changes with respect to its neighbouring positions with perturbation. 
	
\begin{figure}
		\begin{subfigure}[b]{3.5cm}
			\centering
			\includegraphics[width=3.45cm,height=3.45cm]{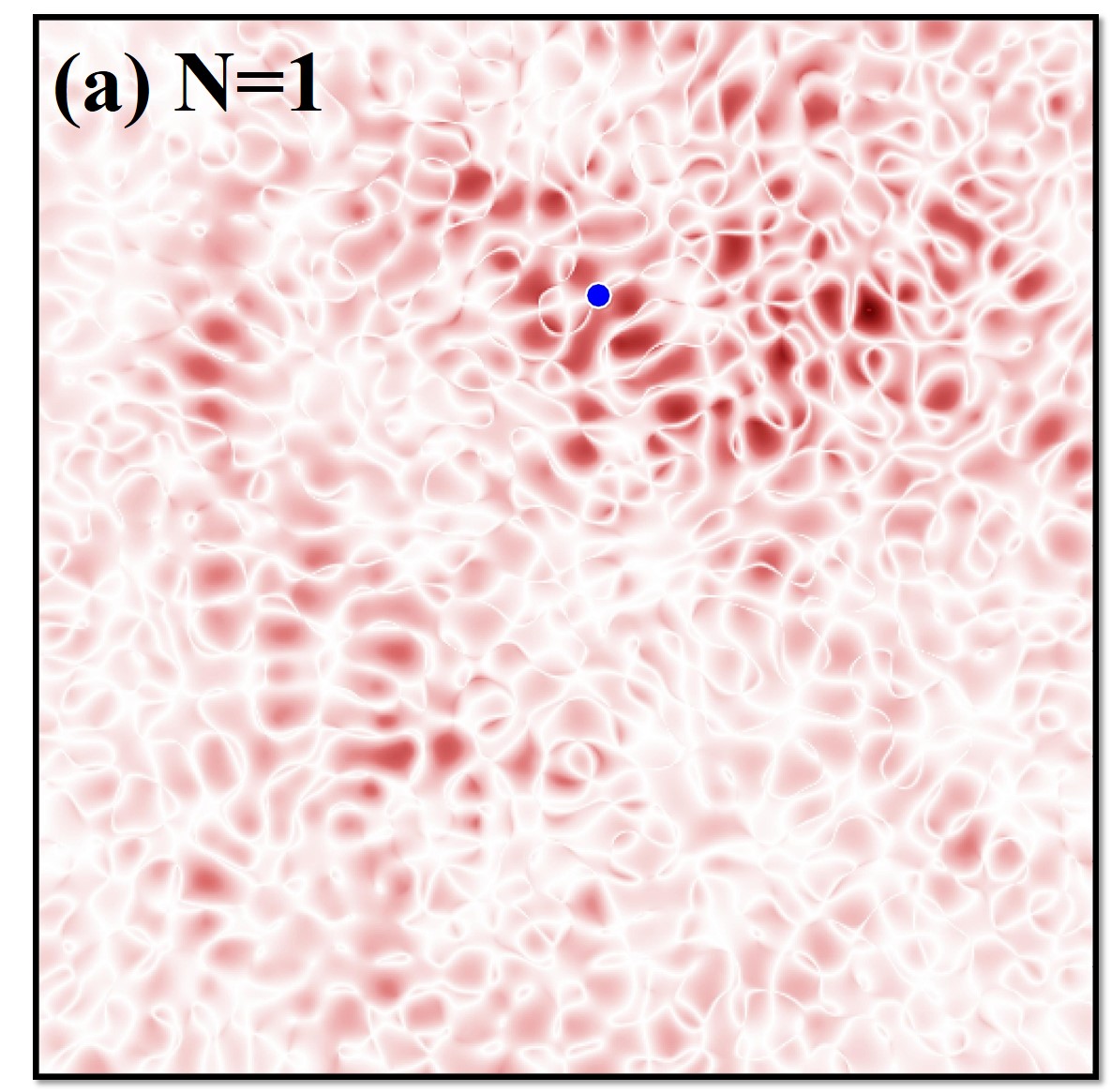} 
			\label{fig4:a} 
		\end{subfigure}
		\begin{subfigure}[b]{3.5cm}
			\centering
			\includegraphics[width=3.45cm,height=3.45cm]{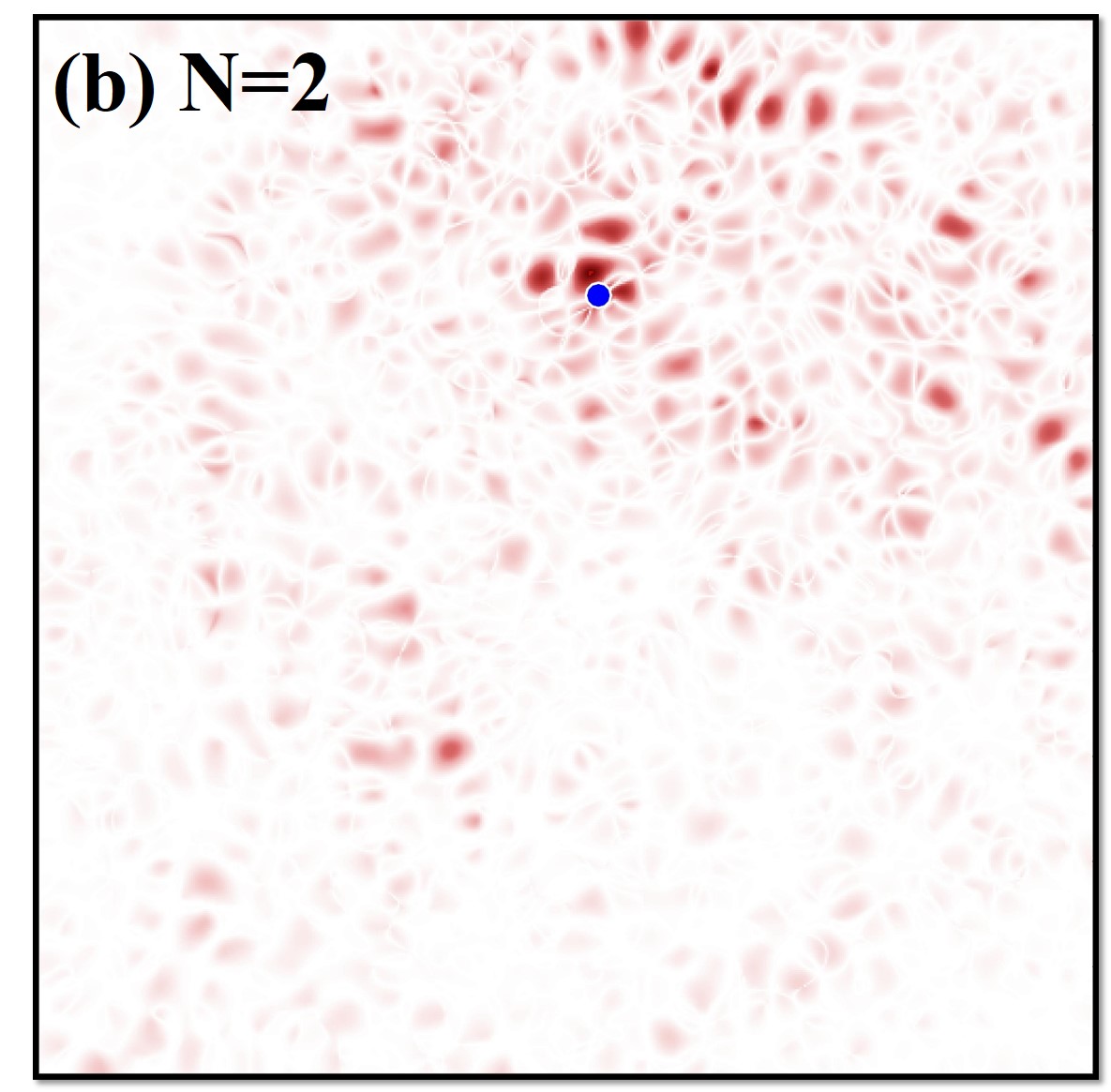} 
			\label{fig4:b} 
		\end{subfigure} 
		\\[-2ex]
		\begin{subfigure}[b]{3.5cm}
			\centering
			\includegraphics[width=3.45cm,height=3.45cm]{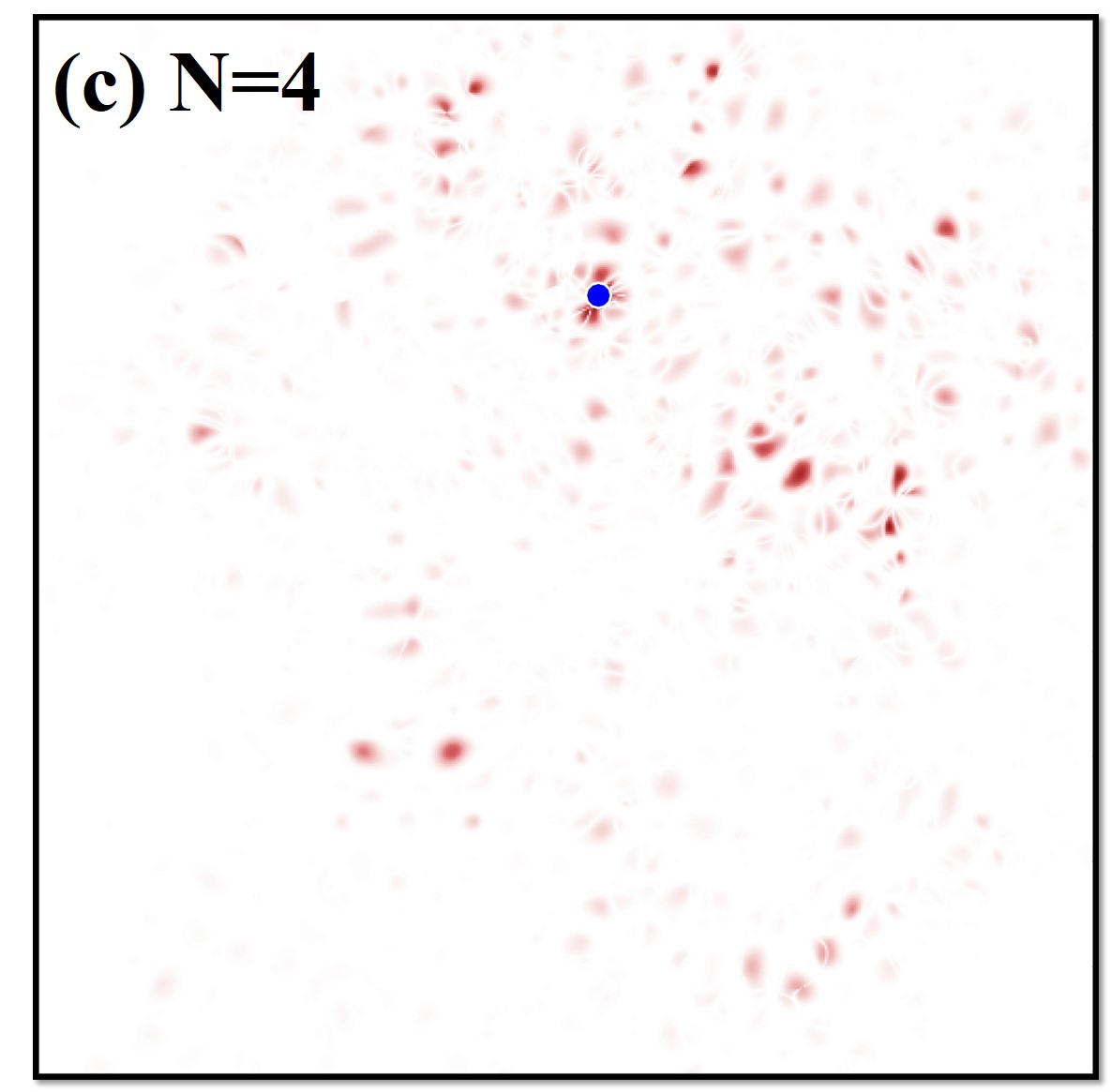} 
			\label{fig4:c} 
		\end{subfigure}
		\begin{subfigure}[b]{3.5cm}
			\centering
			\includegraphics[width=3.45cm,height=3.45cm]{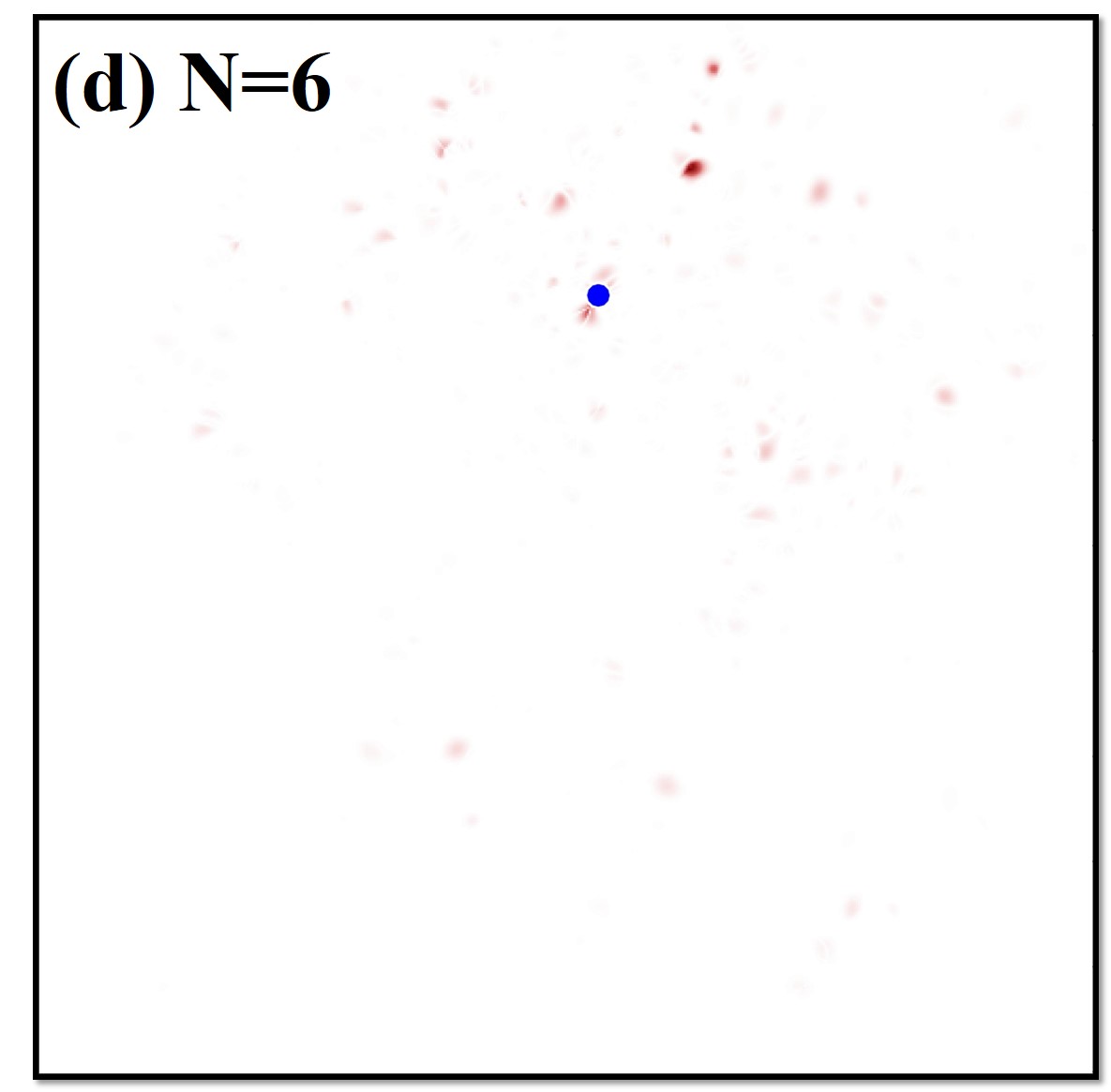} 
			\label{fig4:d} 
		\end{subfigure} 
		\\[-2ex]
		\begin{subfigure}[b]{3.5cm}
			\centering
			\includegraphics[width=3.45cm,height=3.45cm]{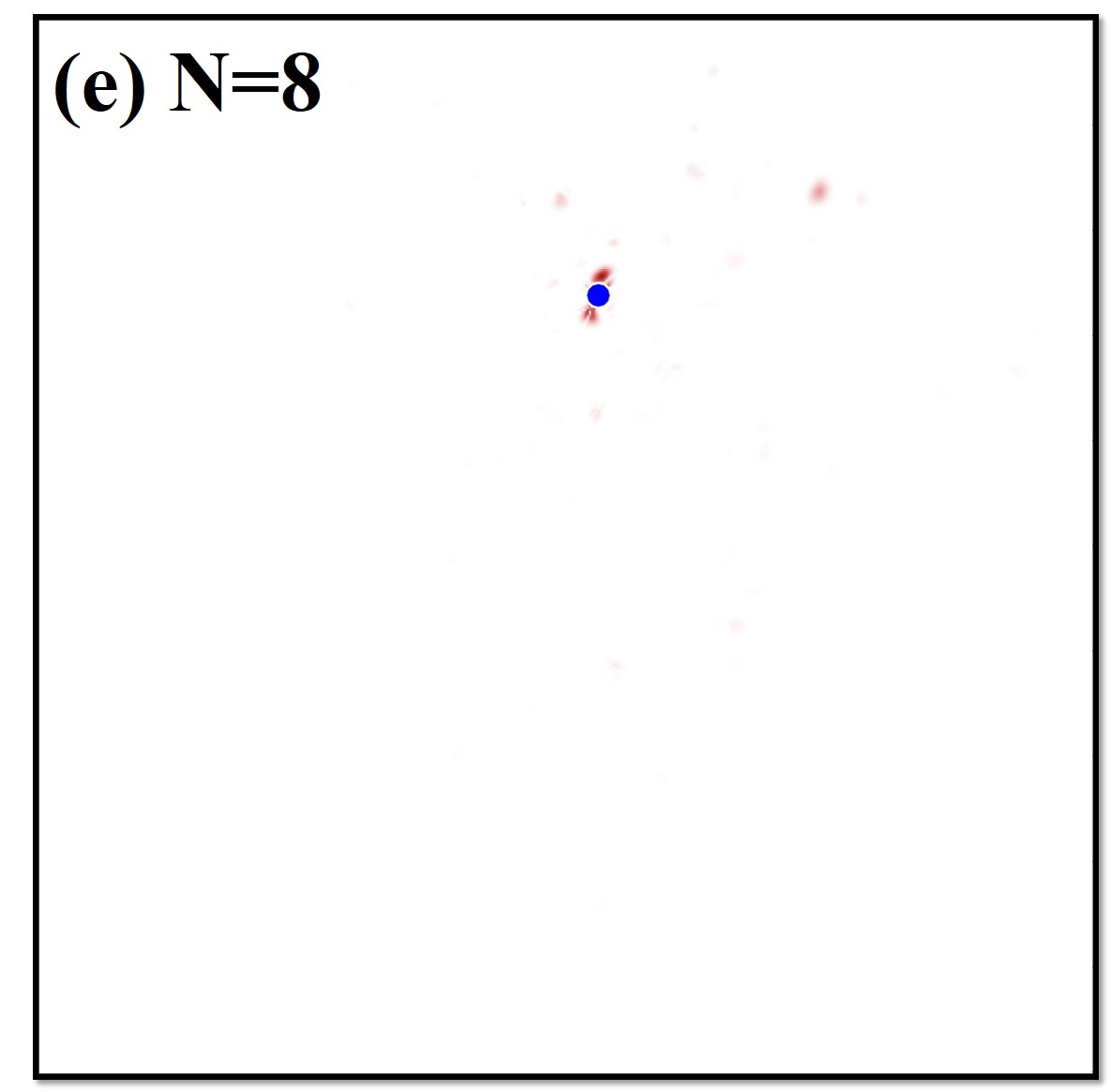} 
			\label{fig4:e} 
		\end{subfigure} 
		\begin{subfigure}[b]{3.5cm}
			\centering
			\includegraphics[width=3.45cm,height=3.45cm]{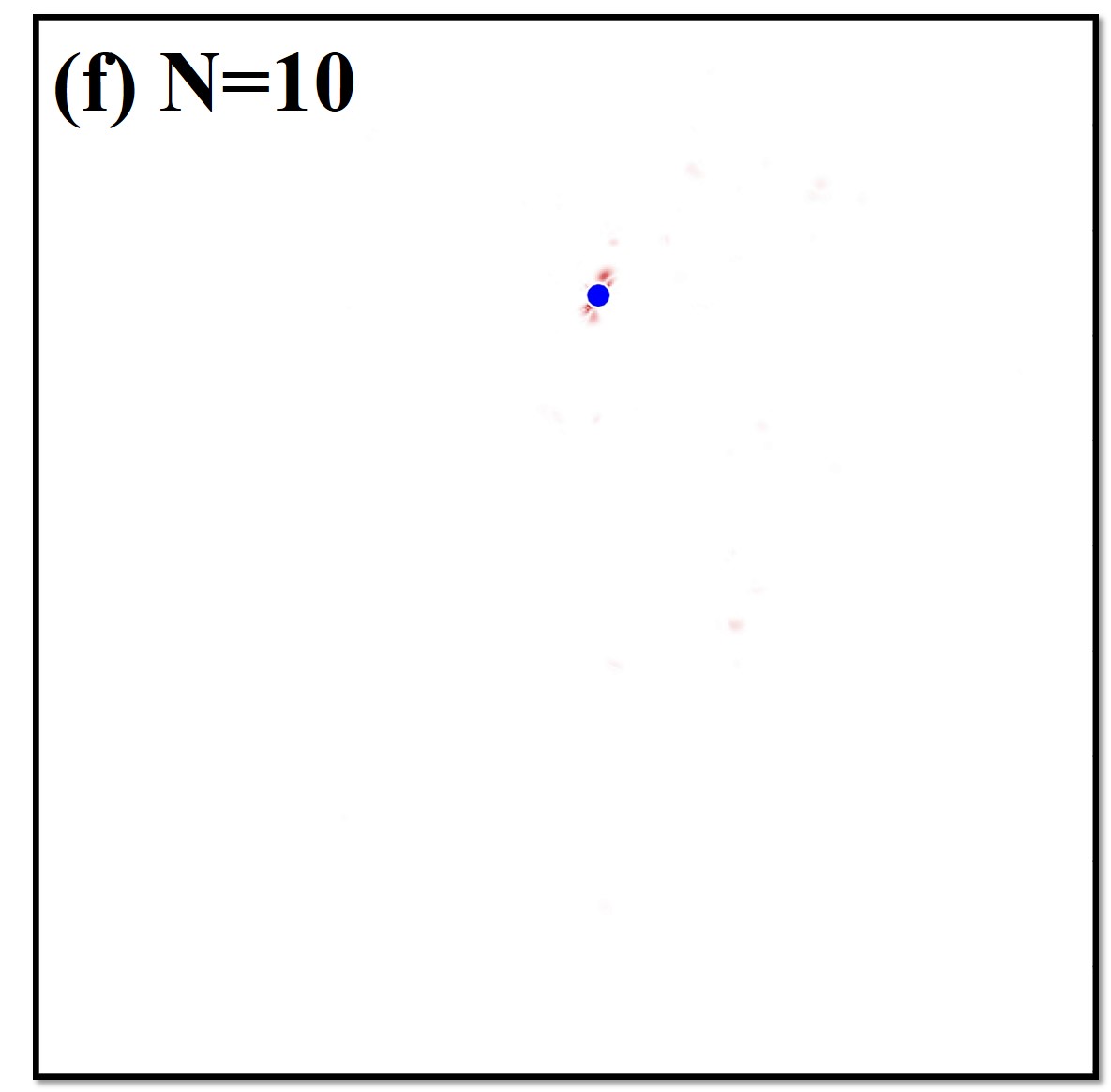} 
			\label{fig4:f} 
		\end{subfigure} 
		\captionsetup{justification=raggedright}
		\caption{The maps generated to locate the position of perturbed particle with the help of TP. The TP mapped regions when number of modes, (a) N=1, (b) N=2, (c) N=4, (d) N=6, (e) N=8, and, (f) N=10 are considered for the particle perturbed by $ 10 \hspace{0.1 cm} nm$ at location 1 (Fig. \ref{fig3}). The perturbed particle is marked in blue.}
		\label{fig4} 
	\end{figure}

	In Fig. \ref{fig4}(a), $TP$ shows the impact of perturbation on a single mode of the system. This contrasts with Fig. \ref{fig2} that shows that modes are rather preserved after perturbation. Thus, for N = 1,  $TP$ quantifies the impact of perturbation on a single mode and implies that a single mode is indeed sensitive to any small change in the system, but it fails to identify the location of perturbation. Further, when two modes are considered, a significant reduction in the region mapped by $TP$ is observed as shown in Fig. \ref{fig4}(b). As the number of modes considered to evaluate $TP$ is increased further, the $TP$ mapped region reduces and concentrates around the perturbed particle as shown in Figs. \ref{fig4} (c-f). Thus, $TP$ provides a way to locate the perturbed particle. The accuracy of the localization of the perturbation increases with the number of modes considered to evaluate $TP$. Here, we were limited to ten modes, but by considering more modes (larger system, larger spectral range) the localization can be improved further.
	
\begin{figure}
		\begin{subfigure}[b]{3.5cm}
			\centering
			\includegraphics[width=3.45cm,height=3.45cm]{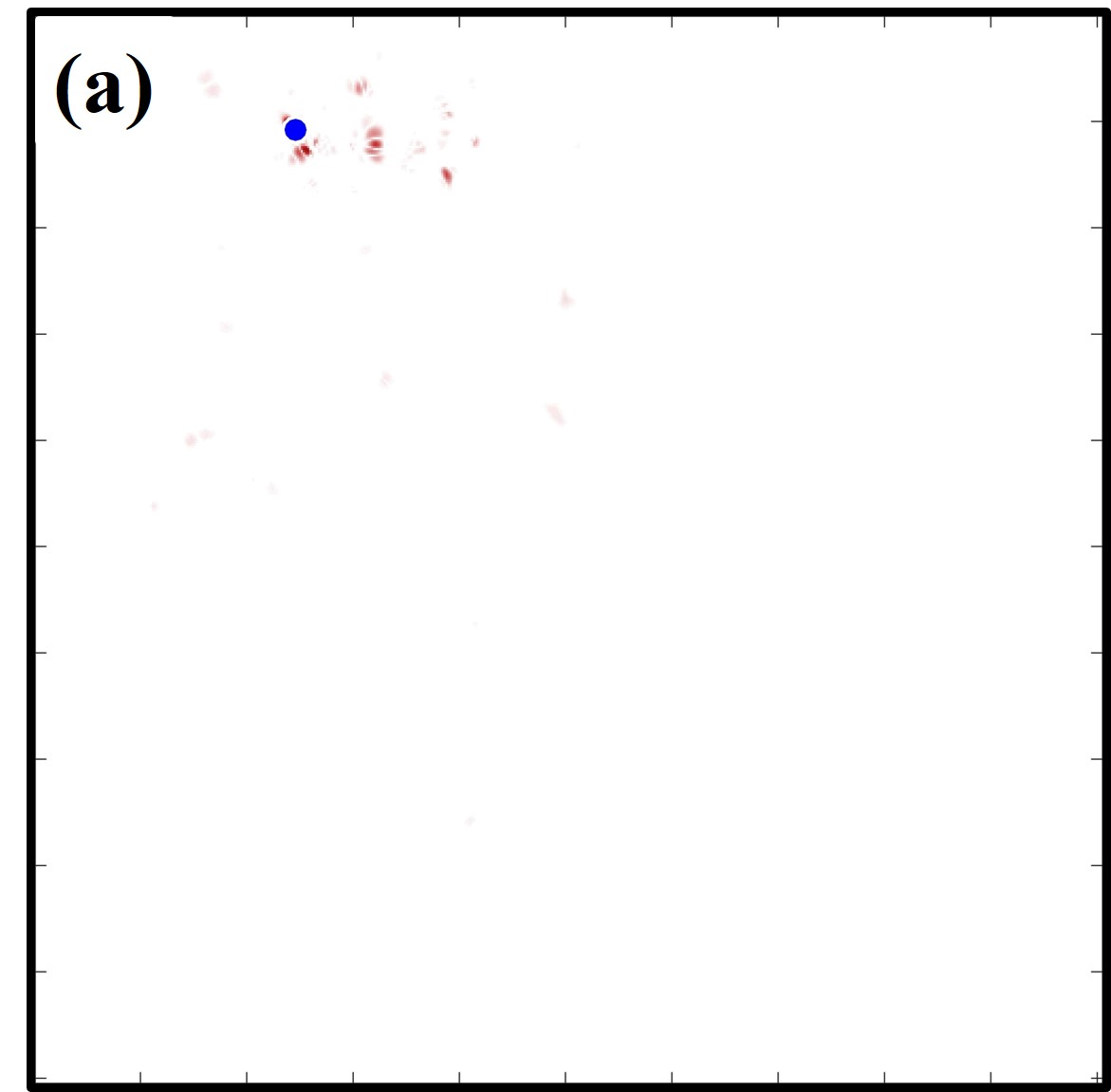} 
			\label{fig5:a} 
		\end{subfigure}
		\begin{subfigure}[b]{3.5cm}
			\centering
			\includegraphics[width=3.45cm,height=3.45cm]{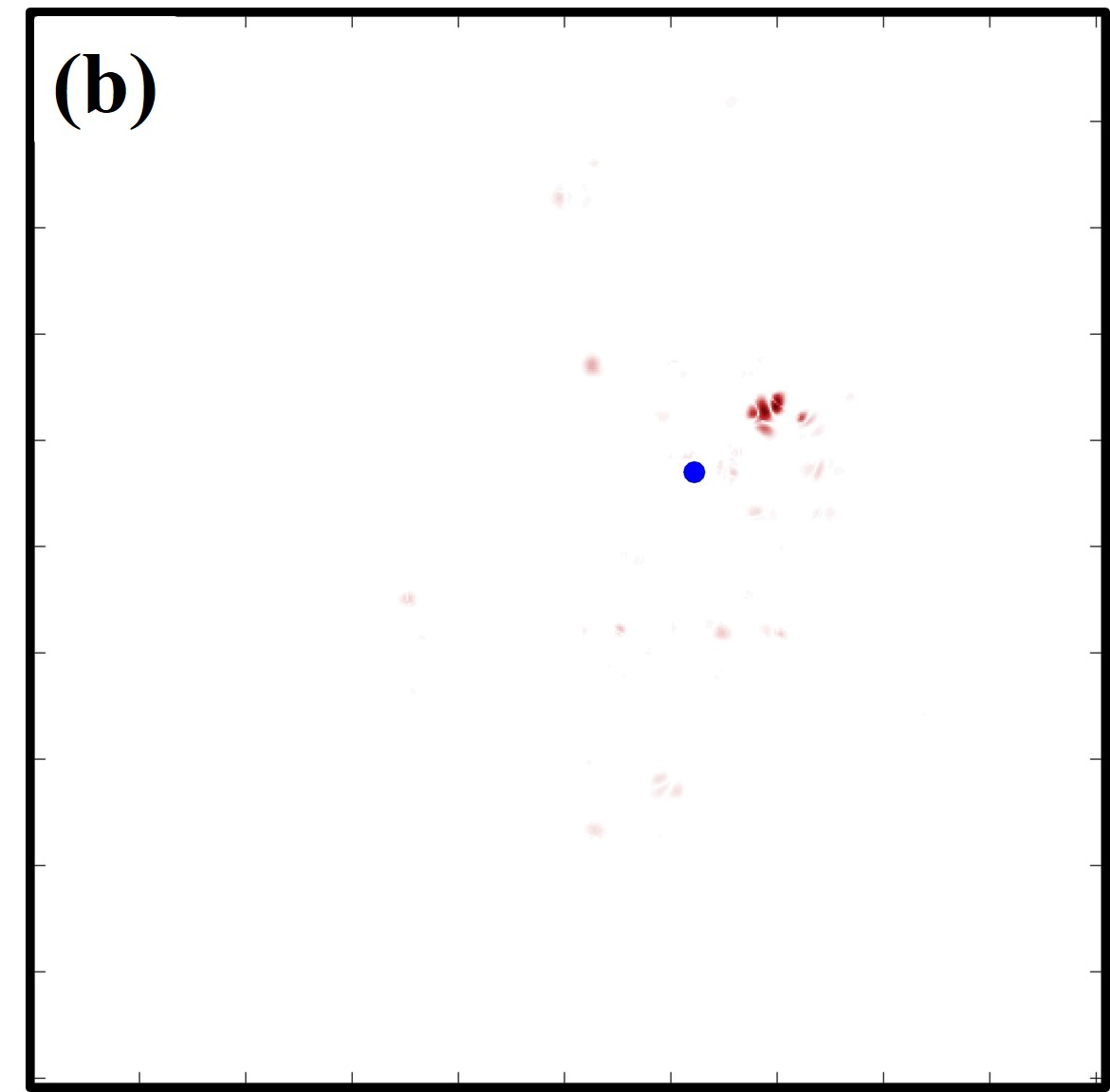} 
			\label{fig5:b} 
		\end{subfigure}
		\\[-2ex]
		\begin{subfigure}[b]{3.5cm}
			\centering
			\includegraphics[width=3.45cm,height=3.45cm]{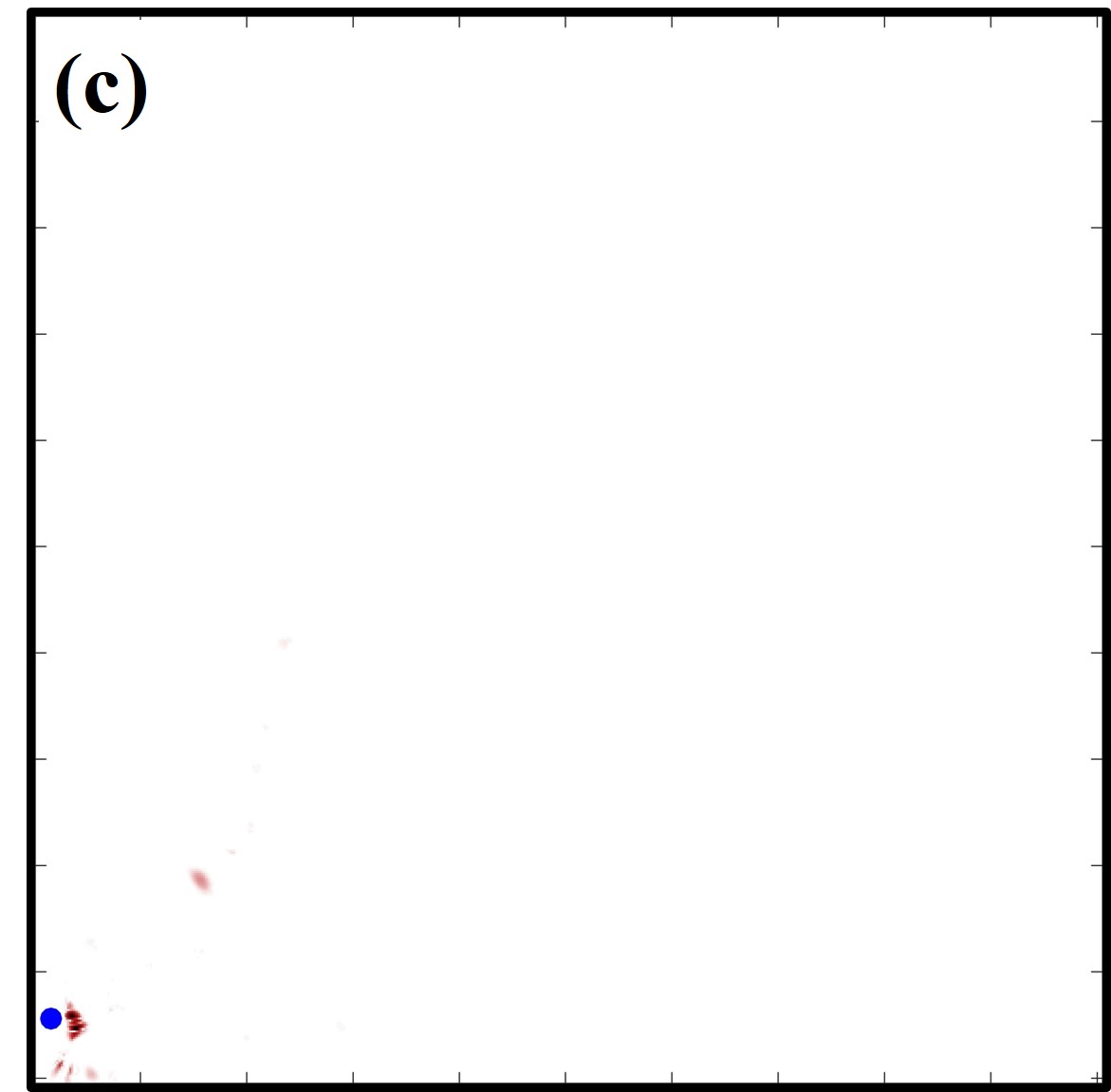} 
			\label{fig5:c} 
		\end{subfigure}
		\begin{subfigure}[b]{3.5cm}
			\centering
			\includegraphics[width=3.45cm,height=3.45cm]{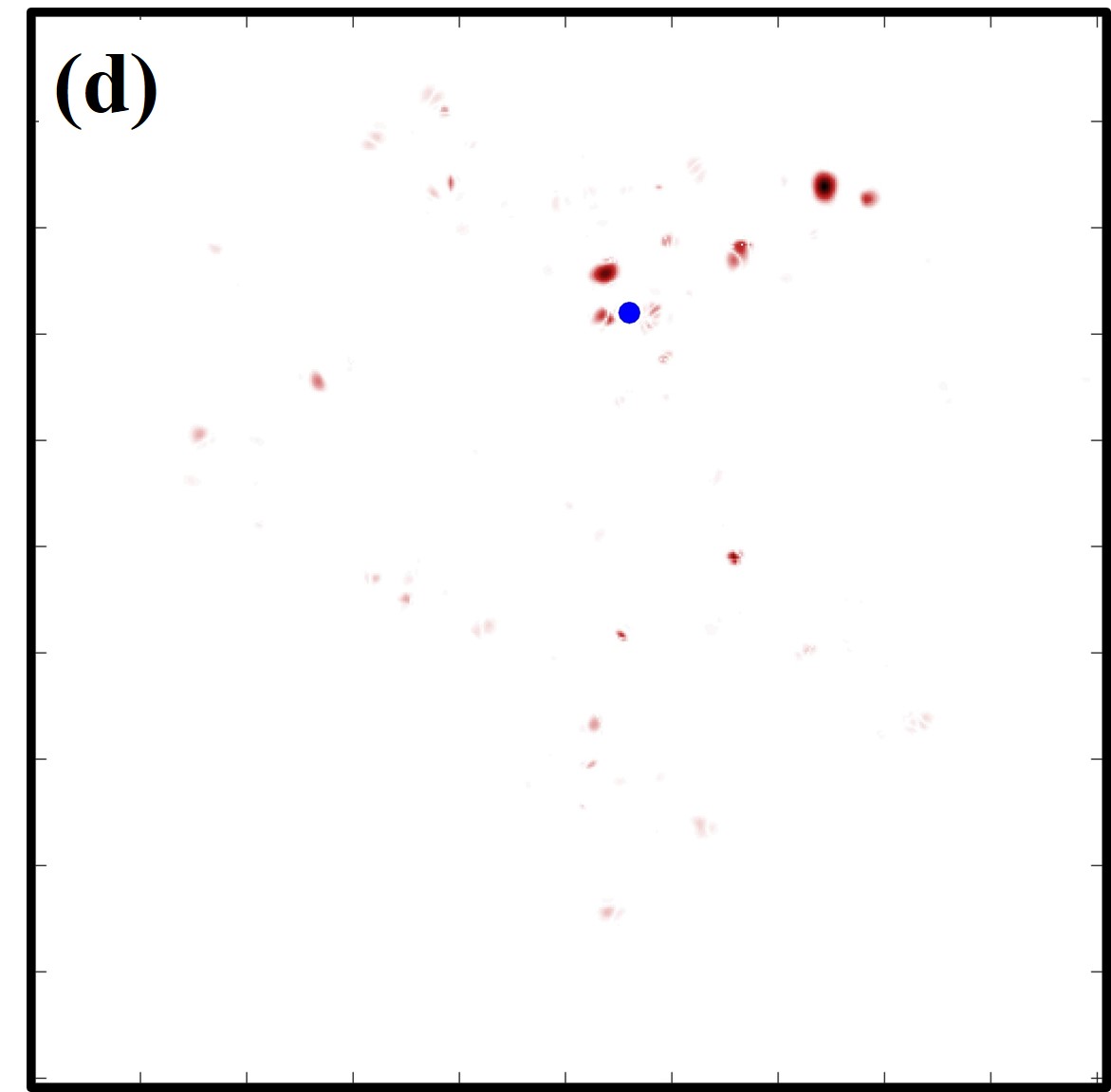} 
			\label{fig5:d} 
		\end{subfigure} 
		\\[-1ex]
		\begin{subfigure}[b]{7 cm} 
			\centering
			\includegraphics[width=6.5cm, height=4.5cm]{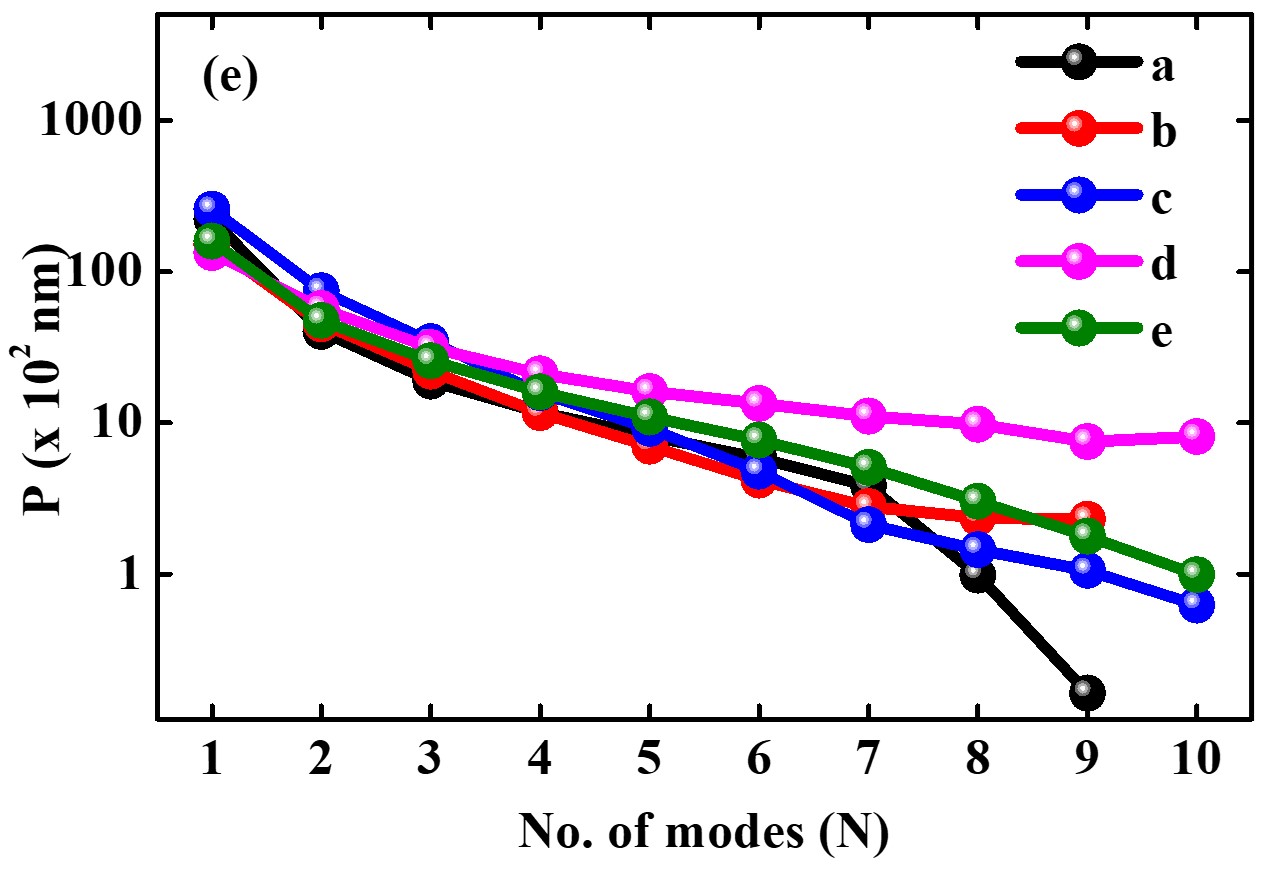} 
			\label{fig4:g}
		\end{subfigure}
		\captionsetup{justification=raggedright}
		\caption{(a-d) The maps generated with the help of $TP$ to locate the position of a particle perturbed by $ 10 \hspace{0.1 cm}nm$ at different locations in  the system. The perturbed particle is marked in blue. (e) The proximity parameter, $P$ as a function of number of modes (N), for systems in (a-d), and for the system in Fig. \ref{fig4} (marked as (e)).}
		\label{fig5} 
	\end{figure}

	Next, to understand the consistency of the proposed approach to locate perturbation, a single particle was displaced at different locations in the system along arbitrary directions by $10 \hspace{0.1 cm}nm$. Figs. \ref{fig5} (a-d), show how the accuracy of the localization of the defect fluctuates from place to place in the system, when ten modes are considered to evaluate $TP$. It is observed that the perturbed particle lies within the mapped region for the perturbation at different locations in the system.
	
	In order to quantify the accuracy to which the location of perturbation has been identified, a proximity parameter has been calculated as a function of number of modes considered. The proximity parameter, $P$ is defined as the root mean square of the distance of the perturbed particle from each point in the tracking parameter mapped region having a $TP$ value of atleast $\frac{1}{e}$ times the maximum TP value. The value of $P$ is defined as follows,
	
	\begin{equation} \label{3}
		P = \sqrt{\frac{1}{n}\sum_{i} d^{2}_i }, \hspace{0.5 cm}
		d_{i}=\sqrt{(x_{p}-x_i)^{2}+(y_{p}-y_i)^{2}}
	\end{equation}
	
	Here, $d_i$ is the distance of perturbed particle at $(x_p,y_p)$ from the location $(x_i,y_i)$ in the region mapped with $TP$. 
	
	Fig. \ref{fig5}(e) shows how  the value of $P$ changes when different number of modes are considered. The value of $P$ for systems in Figs. \ref{fig5} (a-d) are marked as (a-d) in Fig. \ref{fig5} (e). The plot (e) in Fig. \ref{fig5} (e) corresponds to the system perturbed at location 1 in Fig. \ref{fig3}, for which the variation of $TP$ for different number of modes considered is shown in Fig. \ref{fig4}. In Fig. \ref{fig5}(e), it is observed that as the number of modes considered increases, the proximity parameter value decreases, i.e. the region mapped with $TP$ gives a more accurate estimate of the location of perturbation.
			\begin{figure}
		\begin{subfigure}[b]{3.5cm}
			\centering
			\includegraphics[width=3.4cm]{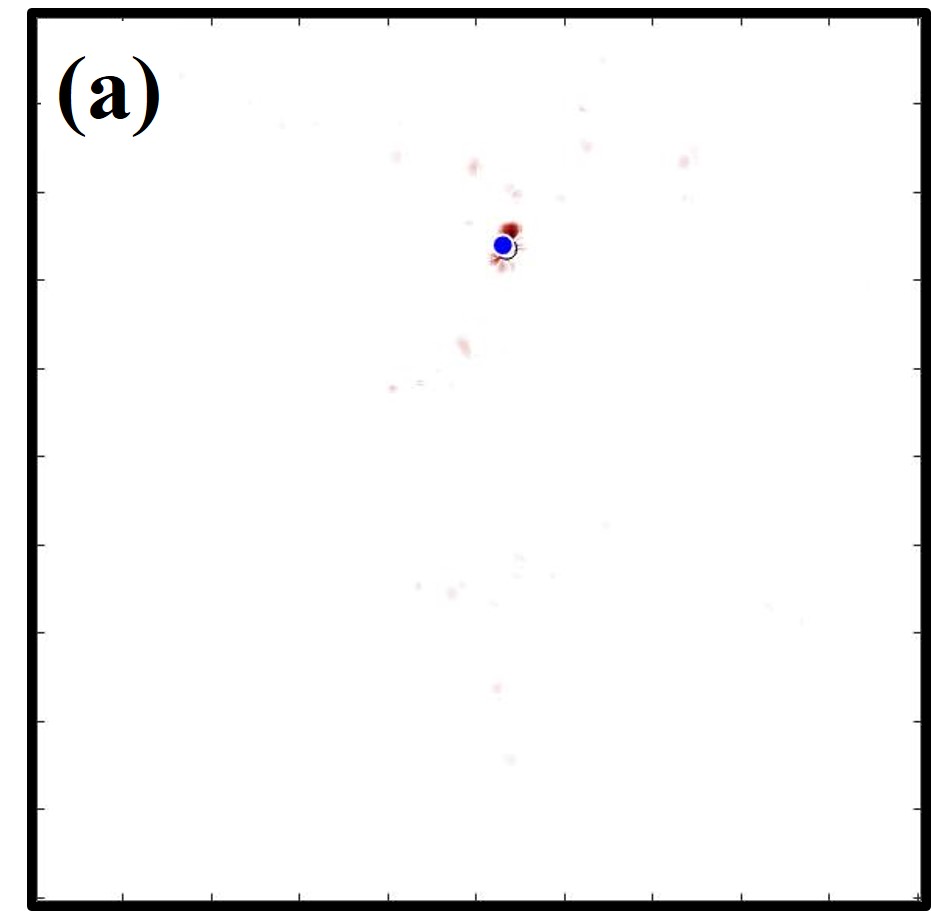} 
			\label{fig6:a} 
		\end{subfigure}
		\begin{subfigure}[b]{3.5cm}
			\centering
			\includegraphics[width=3.4cm]{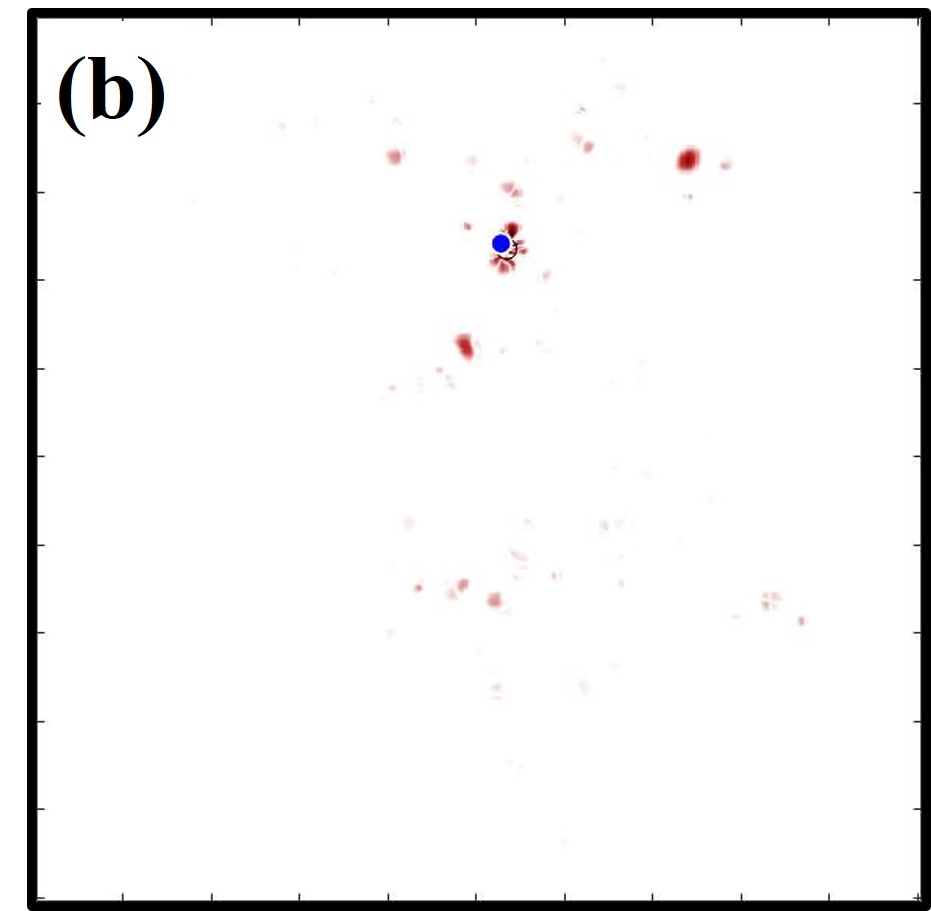} 
			\label{fig6:b} 
		\end{subfigure}
		\begin{subfigure}[b]{3.5cm}
			\centering
			\includegraphics[width=3.4cm]{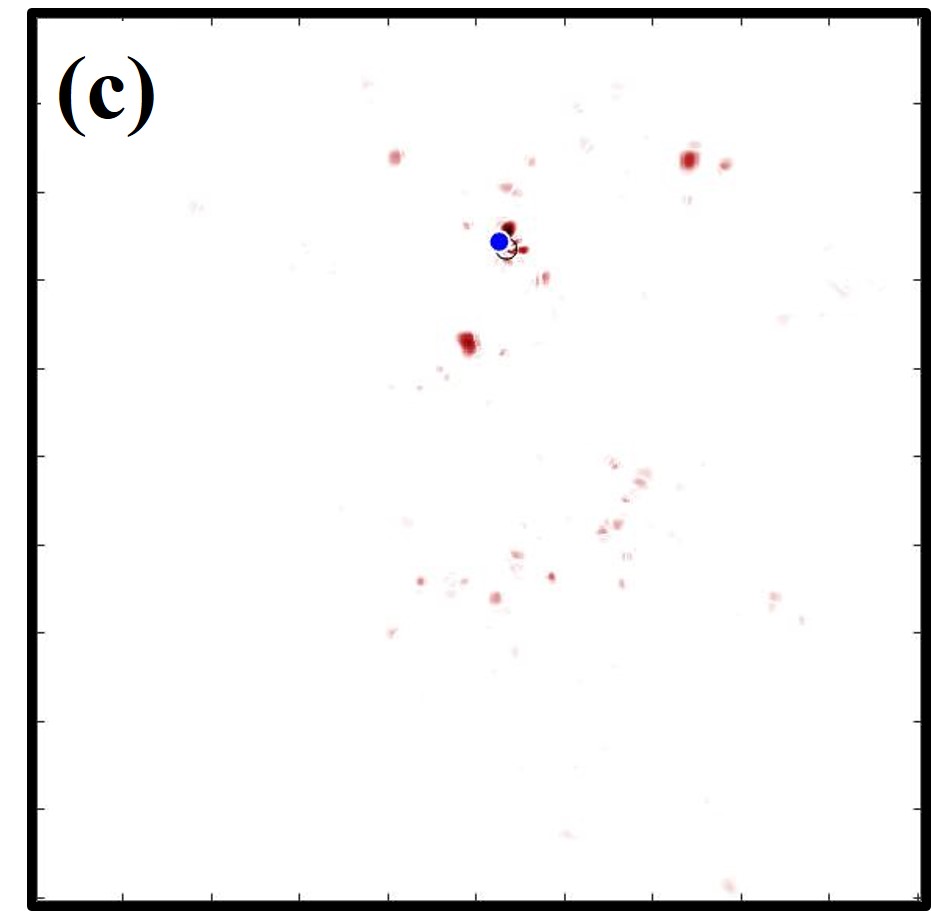} 
			\label{fig6:c} 
		\end{subfigure}
		\begin{subfigure}[b]{3.5cm}
			\centering
			\includegraphics[width=3.4cm]{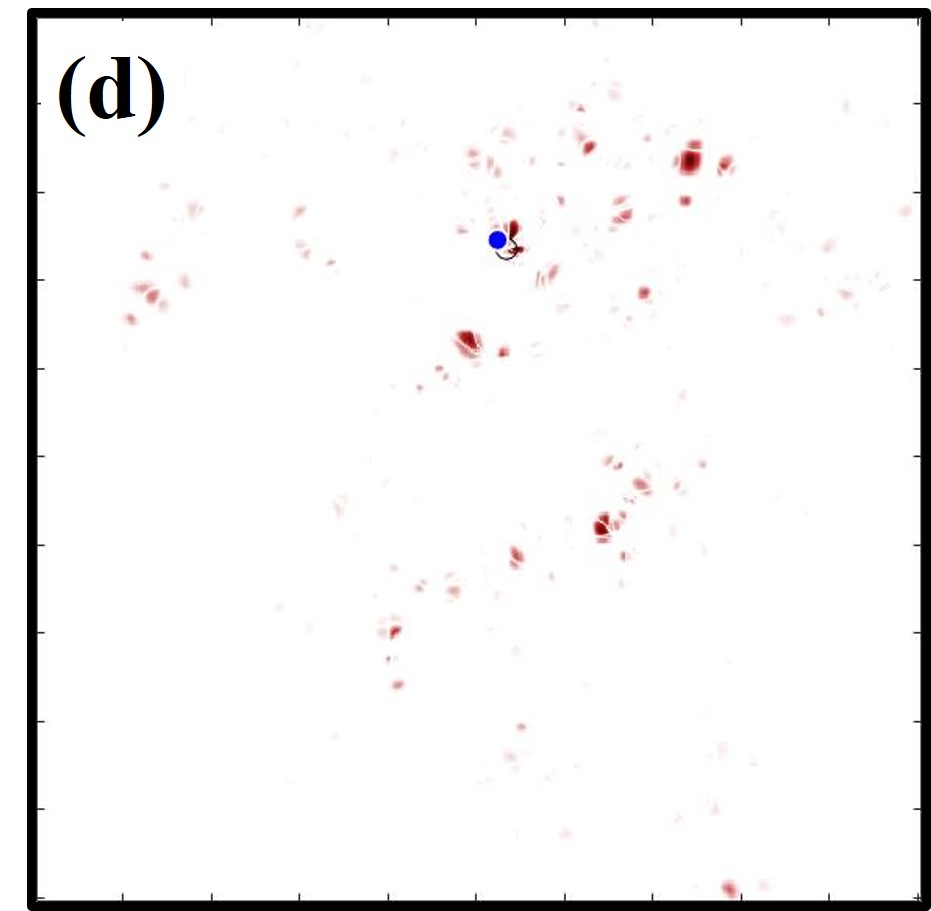} 
			\label{fig6:d} 
		\end{subfigure} 
		\\[2 ex]
		\begin{subfigure}[b]{7 cm}
			\centering
			\includegraphics[width=6.4 cm, height=4.5cm]{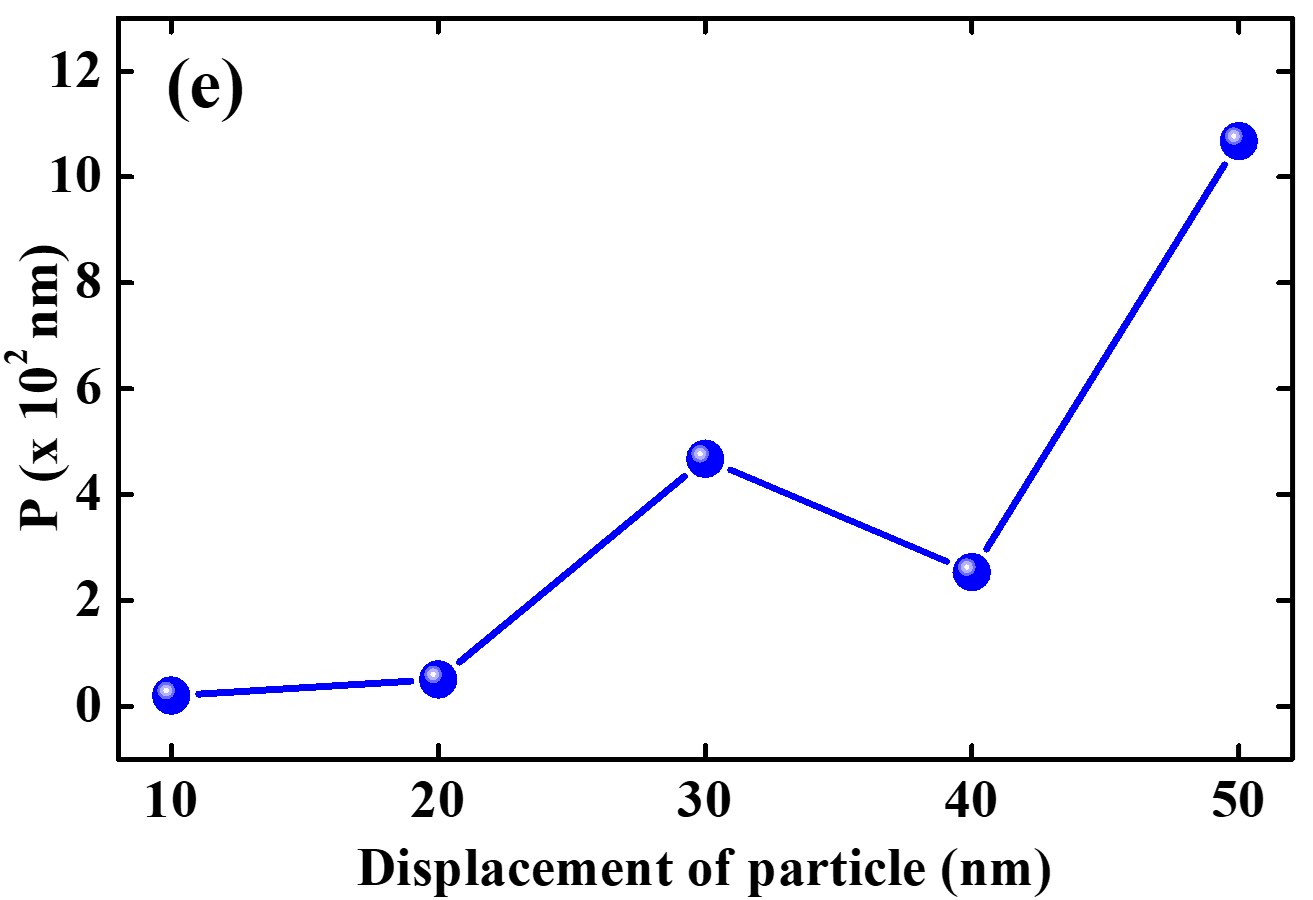}
			\label{fig6:e} 
		\end{subfigure} 
		\captionsetup{justification=raggedright}
		\caption{The maps generated with the help of $TP$ to locate the position of a particle perturbed by (a) $ 20 \hspace{0.1 cm}nm$, (b)$ 30 \hspace{0.1 cm}nm$, (c)$ 40 \hspace{0.1 cm}nm$ and (d)$ 50 \hspace{0.1 cm}nm$ at location 1 in the system. The perturbed particle is marked in blue. (e) The proximity parameter for different values of displacements in (a-d).}
		\label{fig6} 
	\end{figure}
	\FloatBarrier
	To investigate the applicability of the proposed method for larger displacements, the particle perturbed at location 1 in Fig. \ref{fig3} is subjected to  displacements of $20 \hspace{0.1 cm}nm - 50\hspace{0.1 cm}nm $, along an arbitrary direction. It is observed that, for larger displacements, the changes in the mode locations and their corresponding spatial field distributions become more prominent. The region mapped with the help of $TP$ and the displaced particle (marked in blue) are shown in Figs. \ref{fig6} (a-d) when a single particle is perturbed by (a) $20 \hspace{0.1 cm} nm$, (b) $30 \hspace{0.1 cm} nm$, (c) $ 40 \hspace{0.1 cm} nm$ and (d)  $50 \hspace{0.1 cm} nm$.
	 It is observed that as the displacement increases, the mapped regions becomes wider. This result is supported by  the proximity parameter values in Fig. \ref{fig6} (e). Effectively, the tracking parameter can be applicable to accurately identify single-step displacements of $\sim 20 \hspace{0.1cm} nm$ or smaller. Larger displacements can also be accurately mapped if they are carried out in multiple steps of $\sim 20 \hspace{0.1cm} nm$.

	\section{Conclusion}	
	In summary, a numerical study on the detection and localization of nanoscale perturbation in a 2D strongly scattering active disordered system has been presented. The modes and the corresponding spatial intensity distributions have been calculated by solving Maxwell’s equation combined with rate equations for a four level atomic system. A tracking parameter has been proposed to identify the region of nanoscale perturbation. It is shown that the tracking parameter can map the regions of perturbation very well for single nanoscale perturbations. The results presented in this paper demonstrate that nanoscale perturbations in a system can be tracked if the spatial field distribution of the modes before and after the perturbation are known. Thus, RLs have been proposed as a tool to track minute changes taking place in a disordered system. As of now, the tracking parameter can be evaluated with the help of tailored pump intensity profiles used for selective excitation of modes. It was demonstrated recently that localized modes of 1D random laser can be selected and mapped individually \cite{kumar2022investigation}. Our method can be easily tested in this system. It can prove to be useful in biomedical applications to track minute growth of tumor in cells. The imaging methods such as X-ray, CT scan etc. can be used to locate the region of diagnosis and then detailed monitoring of tumor can be carried out with the proposed method.
	
	\begin{acknowledgments}
		The authors acknowledge Jonathan Andreasen, Georgia Tech Research Institute and Anirban Sarkar, National Institute of Technology, Calicut for fruitful discussions and help in computation. We acknowledge support from Science and Engineering Research Board (CRG/2020/002650) sponsored project. DST-FIST facility, Department of Physics, IIT Kharagpur is acknowledged for computational support. We acknowledge National Supercomputing Mission (NSM) for providing computing resources of ‘PARAM Shakti’ at IIT Kharagpur, which is implemented by C-DAC and supported by the Ministry of Electronics and Information Technology (MeitY), Department of Science and Technology (DST), Government of India. Israel Science Foundation (Grants No. 1871/15, 2074/15 and 2630/20), the United States-Israel Binational Science Foundation NSF/BSF (Grant No. 2015694 and Grant No 2021811) are acknowledged.
	\end{acknowledgments}

		\bibliography{manuscript}
		
	\end{document}